\documentclass[fleqn,usenatbib, useAMS, a4paper]{mnras}

\usepackage{savesym}
\savesymbol{tablenum}
\usepackage{siunitx}
\restoresymbol{SIX}{tablenum}

\usepackage{newtxtext}
\usepackage[varg,varvw,smallerops]{newtxmath}

\usepackage[T1]{fontenc}
\usepackage{ae,aecompl}

\usepackage{graphicx}	

\usepackage{amsmath}	
\usepackage{amssymb}	
\usepackage{multicol}
\usepackage{enumerate}          
\usepackage{xcolor}

\usepackage[spanish,es-minimal,english]{babel}

\usepackage{booktabs}
\usepackage{array}   
\newcolumntype{L}{>{$}l<{$}} 
\newcolumntype{R}{>{$}r<{$}} 
\newcolumntype{C}{>{$}c<{$}}

\hypersetup{hidelinks=True}

\usepackage{upgreek}
\usepackage{placeins}
\definecolor{NEWcolor}{rgb}{0.7,0.2,0.1}

\definecolor{MAYBEcolor}{rgb}{0.07,0.92,0.16}

\newcommand\pos{\ensuremath{_{\mathrm{pos}}}}

\newcommand\noise{\ensuremath{_{\text{noise}}}}
\newcommand\obs{\ensuremath{_{\mathrm{obs}}}}
\newcommand\model{\ensuremath{_{\mathrm{mod}}}}

\newcommand\ha{\ensuremath{\text{H}\upalpha}}

\newcommand\Wav[1]{\ensuremath{\lambda #1}}

\newcommand\xx{\ensuremath{\boldsymbol{x}}}
\newcommand\PM[2]{\ensuremath{\substack{+#1\\-#2}}}

\newcommand\ariii{[\ion{Ar}{III}] \Wav{7136}}
\newcommand\hii{\ion{H}{II}}
\newcommand\halpha{\ha}
\newcommand\n{[\ion{N}{II}] \Wav{6584}}
\newcommand\oi{[\ion{O}{I}] \Wav{6300}}
\newcommand\oiii{[\ion{O}{III}] \Wav{5007}}
\newcommand\sii{[\ion{S}{II}] \Wav{6731}}
\newcommand\siii{[\ion{S}{III}] \Wav{9069}}
\newcommand\vcentroid{\ensuremath{V_{\text{c}}}}

\newcommand\binsize{\ensuremath{\Delta x_{\text{bin}}}}

\newlength\SFwidth
\newcommand\FITtwograph[2]{%
  \includegraphics[width=\SFwidth]{figures/sf-emcee-#1}
  &  \includegraphics[width=\SFwidth]{figures/corner-emcee-#1}
}

\newcommand\fitfigg[2]{%
  \begin{tabular}{@{}ll@{}}
    (a)& (b)\\
    \FITtwograph{#1}{#2}
  \end{tabular}%
}

\newcommand\fitfigggg[4]{%
  \begin{tabular}{@{}ll@{}}
    (a)& (b)\\
    \FITtwograph{#1}{#2}\\
    (c)& (d)\\
    \FITtwograph{#3}{#4}\\
  \end{tabular}%
}

\title[Turbulent velocity statistics from IFUs]{Recovering turbulent velocity statistics from noisy integral-field spectroscopy}

\author[J. García-Vázquez et al.]{
  J. García-Vázquez\textsuperscript{1}\thanks{j.garcia@irya.unam.mx},
  William J. Henney\textsuperscript{1}\thanks{w.henney@irya.unam.mx},
  and S. Jane Arthur\textsuperscript{1}
  \\
  \textsuperscript{1}Instituto de Radioastronomía y Astrofísica,
    Universidad Nacional Autónoma de México,
    Antigua Carretera a Pátzcuaro 8701,
    58089 Morelia, Michoacán, Mexico\\
}

\date{Accepted XXX. Received YYY; in original form ZZZ}

\pubyear{2026}

\begin{document}
\label{firstpage}
\pagerange{\pageref{firstpage}--\pageref{lastpage}}
\maketitle

\begin{abstract}
Turbulence plays a critical role in the evolution of \hii{}
regions, yet recovering its statistical properties from observations
remains challenging, particularly when centroid velocities are
affected by instrumental noise. We present a methodology for
recovering the second-order velocity structure function from
intermediate spectral-resolution integral-field spectroscopy and
apply it to VLT MUSE observations of the Orion Nebula. The
plane-of-sky velocity field is characterized by fitting a simple
parametric model to the observed structure function, allowing the
turbulent velocity variance, correlation length, power-law slope,
and noise level to be estimated simultaneously. We investigate the
trade-off between instrumental noise and spatial resolution through
systematic spatial binning and validate the recovered turbulent
parameters against previously analyzed high spectral-resolution
KPNO echelle observations. We find that spatial binning improves
the empirical structure function provided that the bin size remains
below approximately \(0.05\,r_0\), while the fitted model recovers
consistent turbulent parameters across all binning levels. The MUSE
results agree closely with those from the KPNO data, demonstrating
that intermediate-resolution integral-field spectroscopy can recover
reliable turbulence statistics despite its relatively poor velocity
resolution. We also identify modest but systematic differences in
the turbulent properties traced by emission lines of different
ionization potential, reflecting the geometric and ionization
structure of the nebula. Our methodology provides a robust framework
for extracting turbulent velocity statistics from noisy
integral-field spectroscopic observations.
\end{abstract}

\begin{keywords}
HII regions -- ISM: kinematics and dynamics -- turbulence 
\end{keywords}



\section{Introduction}

The process of star formation begins with the gravitational collapse of molecular clouds, followed by fragmentation into dense cores where stars form in clusters \citep{2003RPPh...66.1651L, 2007ARA&A..45..565M}. 
Massive O- and B-type stars emerge within these stellar groups, and their intense ionizing radiation and stellar winds inject substantial energy and momentum into the surrounding gas, profoundly influencing the structure and dynamics of the the recently created ionized regions \citep{2006ApJ...647..397M, 2024RMxAC..58....8A}.

\hii\ regions reveal heterogeneous velocity patterns that deviate from simple radial expansion \citep{TT1979,1996AJ....111.2349H,2015A&A...573A..10M,2019MNRAS.487.2200Z}.
Instead, the ionized gas exhibits complex behavior that can be interpreted as turbulence. 
Within this framework, \hii\ regions offer ideal laboratories for studying turbulent dynamics in the astrophysical context \citep{1949ApJ...110..329C}.

Turbulence in \hii\ regions is typically first characterized through the amplitude of velocity fluctuations. 
This can be studied using velocity dispersion along the line-of-sight (LOS) or through the variance of centroid velocities projected onto the plane-of-sky (POS) \citep{1988A&A...201..199A,arthur2016turbulence}. 
Statistical tools, such as the second-order structure function, enable the recovery of turbulence signatures in the POS such as the power-law $m$ and correlation length \(r_0\), and allow comparisons with theoretical models such as the Kolmogorov theory of incompressible turbulence \citep{von1951methode,1955IAUS....2..131C,munch1958internal,1961MNRAS.122....1F,1970A&A.....8..486L,Roy:1985a,1986ApJ...300..624R,1987ApJ...317..686O, castaneda1988,Mivi1995,1997ApJ...487..163M,Chakraborty:1999a, lagrois2011,arthur2016turbulence,2019arXiv191203543M,garciav23,2025ApJ...986..159R,2026A&A...707A.339R}.
However, recovering these turbulent parameters from observational data is not straightforward.

Given that the structure function characterizes how velocity fluctuations vary with spatial separation, it is essential that the measured scales reflect genuine turbulent motions rather than instrumental effects.
In practice, this condition is difficult to satisfy. 
Ground-based observations are affected by atmospheric seeing, while the smallest measurable velocity fluctuations are limited by the instrument’s spectral resolution and sensitivity. 
In addition, noise and other systematic effects can distort the observed velocity field. 

Note that we use the term ``turbulence'' as a shorthand
for all globally disordered and chaotic motions within an \hii{} region,
even though these motions do not necessarily represent homogeneous hydrodynamic turbulence
in the sense of \citet{kolm1}.
The aim of the present study is to characterize the spatial
statistics of the measured velocity field in the ionized gas,
which inevitably incorporate contributions from photoevaporative flows, ionization-front geometry, other large-scale ordered motions, along with genuine turbulent fluctuations.
The physical origin of these velocity fluctuations, and the relative contributions of photoevaporative flows and turbulence, were examined in detail in earlier papers \citep{Medina:2014a, arthur2016turbulence},
where empirical results were compared with synthetic observations from numerical simulations. See especially Fig.~16 of \citet{arthur2016turbulence},
which summarizes the conceptual scheme of causal relationships.
The present paper has a narrower goal: to assess the methodology used to retrieve the characteristic fluctuation amplitude, correlation length, and power-law slope from observed structure functions.

The Orion Nebula (M~42), located at a distance of
\qty{388(5)}{pc}
\citetext{\(\qty{1}{\arcsecond} \approx \qty{0.0019}{pc}\); \citealp{Kounkel:2017a}},
is the closest massive star-forming region and one of the most studied \hii\ regions. 
Ionized by the O7~V star $\theta^1$\,Ori~C, the nebula exhibits complex physical and kinematic structures, including a rich population of young stars, stellar outflows, and Herbig–Haro objects \citep{1993ApJ...410..696O,2001ARA&A..39...99O,Garcia-Diaz:2007a,2009AJ....137..367O,2019MNRAS.486.3423H,2021MNRAS.502.1703M}. 
Its proximity and brightness make it a prime target for studies using spectroscopic observations \citep{Garcia-Diaz:2008a,2015A&A...582A.114W,2016ApJ...819..136A}.

Several studies have investigated turbulence in the Orion Nebula using statistical tools such as the second-order structure function \citep{von1951methode,1955IAUS....2..131C,munch1958internal,castaneda1988,1992ApJ...387..229O,arthur2016turbulence,2016MNRAS.455.4057M,2019MNRAS.483..704A}. 
However, these analyses have led to significantly different conclusions regarding the nature of the turbulent cascade. 
For example, \citet{arthur2016turbulence}, using Kitt Peak National Observatory (KPNO) echelle observations, reported clear power-law behavior consistent with a turbulent cascade, while \citet{2016MNRAS.455.4057M}, analyzing Very Large Telescope (VLT) Multi Unit Spectroscopic Explorer (MUSE) data, found little evidence of such structure. 
More recent studies using the same VLT MUSE observations \citep{2019MNRAS.483..704A} have obtained intermediate results.

\begin{figure*}
 \centering
 \includegraphics[width=6.5in]{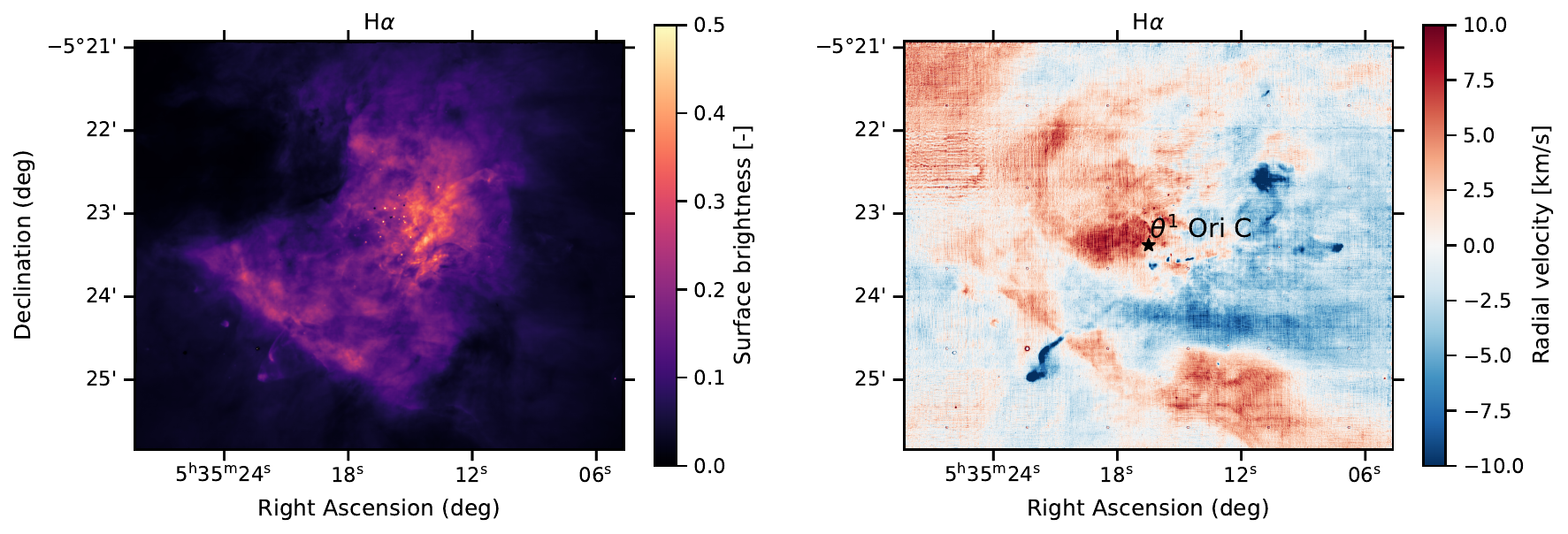}\par
 \caption{
 Two-dimensional maps of the normalized \halpha\ integrated intensity (surface brightness) and centroid velocity in the Orion Nebula. 
 The surface brightness is normalized to the peak intensity in the map, and the velocity scale is shown after subtracting the mean velocity. North is up and East is to the left.
 }
\label{fig:H_I-6563_maps}
\end{figure*}

\begin{figure*}
 \centering
 \includegraphics[width=5in]{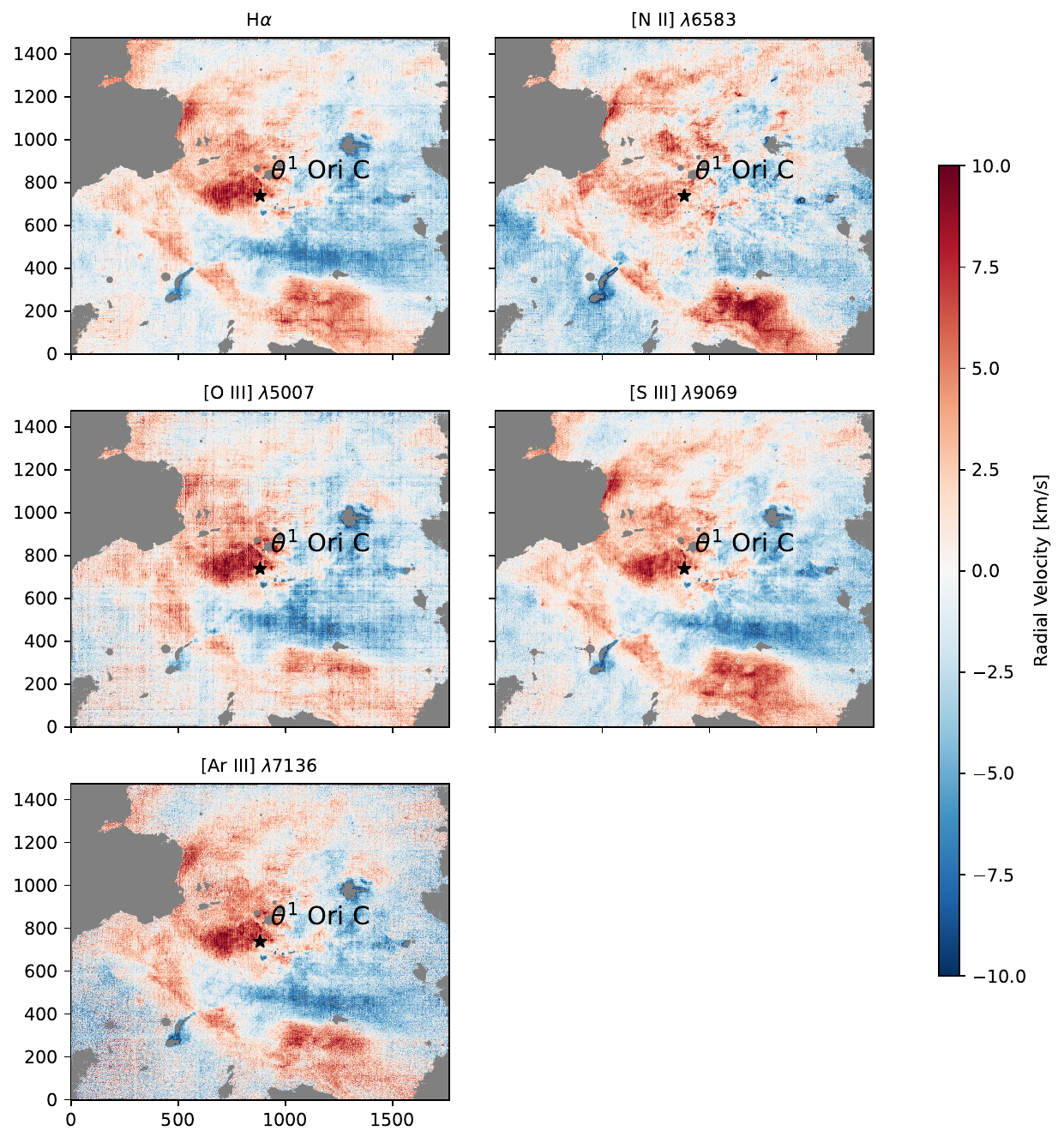}\par
 \caption{
   Two-dimensional velocity maps for different emission lines of the Orion Nebula obtained from VLT MUSE observations,
   with axes marked in native MUSE pixels of size \qty{0.2}{arcsec}. 
   We use the continuum map to identify and exclude pixels associated with stars in the region. 
   Using the \halpha\ emission line as a reference, we further exclude pixels with normalized surface-brightness values below \(0.025\), as well as high-velocity pixels associated with Herbig-Haro objects and proplyds. 
   The same selection criteria are then applied consistently to the remaining emission lines.
   Each emission line is centered with respect to its own mean centroid velocity value.
   Omitted pixels are shown in grey.}
 \label{fig:velocity_maps_shared_cbar_fixed}
\end{figure*}

The KPNO and MUSE datasets differ significantly in both their spectral and spatial sampling properties. 
The KPNO echelle observations have an instrumental spectral resolution of approximately \qty{8}{km.s^{-1}}, whereas the MUSE observations have a spectral resolution of about \qty{120}{km.s^{-1}} at \halpha. 
Conversely, the MUSE observations provide finer spatial sampling, with a pixel scale of \qty{0.2}{arcsec.pixel^{-1}}, compared to \num{1}--\qty{2}{arcsec.pixel^{-1}} for the KPNO data. 
The combination of lower instrumental noise and higher spectral resolution makes the KPNO observations a useful reference for assessing the turbulent parameters recovered from the MUSE data.

The discrepancies in different works are likely driven by differences in methodology, data quality, and the treatment of observational effects such as noise and spatial resolution \citep{garciav23}. 
In particular, instrumental noise can obscure small-scale velocity fluctuations, making the recovery of the turbulent cascade highly sensitive to the analysis technique. This raises a fundamental question: how can the turbulent structure function be reliably recovered from noisy integral field spectroscopic observations? 

More generally, the lack of a standardized methodology for extracting turbulent parameters from observational data makes direct comparisons between studies difficult. 
This fragmentation poses a major challenge for turbulence research, especially in the absence of a complete astrophysical turbulent theory, where observational constraints play a central role.

In \citet{garciav23}, we introduced a heuristic model for the second-order structure function and applied it to \hii\ regions of varying sizes. 
This approach provided a coherent methodology that not only enabled the recovery of turbulent parameters, but also allowed for meaningful comparisons between different regions.

In this work, we address these issues by implementing a consistent methodology to recover the turbulent structure function from noisy observations. 
By combining masking, controlled binning, and model fitting, we aim to reliably extract the turbulent parameters and provide a framework for meaningful comparison between different datasets.
We also extend its use to new observational data and address how to mitigate nuisance parameters that commonly affect the measurement of turbulence in ionized gas.

In Section~\ref{sec:observations}, we describe the VLT MUSE integral-field spectroscopic observations of the Huygens region in the Orion Nebula, which we use as a concrete example to analyze the turbulent velocity field across multiple emission lines: \halpha, \oiii, \n, \siii, and \ariii.
We also describe KPNO echelle observations which we use as reference to compare VLT MUSE results. 
In Section~\ref{sec:met}, we combine a systematic selection procedure with a downsampling strategy to recover the intrinsic second-order structure function of the velocity fluctuations. 
The resulting turbulent parameters are then derived and analyzed in Section~\ref{sec:results}.
In Section~\ref{sec:discussion}, we examine how the recovered turbulent parameters vary between emission lines and datasets, discuss their implications for the turbulent structure of ionized gas and compare with previous results. 
Particular attention is given to the extent to which the turbulent cascade can be recovered from noisy observational data.
Finally, Section~\ref{sec:summary} summarizes our results.

\section{Observations}\label{sec:observations}

The Orion Nebula was observed with MUSE during its commissioning phase on 2014 February 16 \citep{2014Msngr.157...13B}.
A total of \(60 \times 5 \ \text{s}\) exposures were used to build a \(\qty{6}{arcmin} \times \qty{5}{arcmin}\) mosaic, in which each of the 30 distinct pointings was observed twice, employing a 90\textdegree{} dither rotation.
For details of the data reduction see \citet{2015A&A...582A.114W}.
The observations were carried out in the wide field mode with a field-of-view (FOV) of \(\qty{1}{arcmin} \times \qty{1}{arcmin}\), with a the wavelength range of \(4595-  \qty{9366}{\angstrom} \) and a sampling of \(\qty{0.2}{arcsec} \times \qty{0.2}{arcsec} \times \qty{0.85}{\angstrom}  \). 
The data is in the range of \(6300 - \qty{6800}{\angstrom}\), where the spectral resolving power is \(R \simeq 2500\) \citep{2021MNRAS.502.4597H}.
The derived velocity fields were previously analyzed in \citet{2016MNRAS.455.4057M} and \citet{2019MNRAS.483..704A}.

For comparison with the VLT MUSE observations, we use echelle observations of the Orion Nebula obtained with the \qty{4}{m} telescope at Kitt Peak National Observatory (KPNO), covering the \halpha, \oiii, and \n\ emission lines \citep{Doi:2004a}. 
The \sii\ data are taken from supplementary observations obtained with the Manchester Echelle Spectrometer attached to the \qty{2.1}{m} telescope at the Observatorio Astronómico Nacional at San Pedro Mártir (OAN-SPM), Mexico.
The observations map the central Huygens region of the nebula over an area of approximately \(\qty{3}{arcmin} \times \qty{5}{arcmin}\). 
The dataset is composed of 96 north-south oriented long-slit spectra, each spanning \qty{300}{arcsec}, separated by \qty{2}{arcsec}, and observed with a slit width of \qty{0.8}{arcsec}. 
The resulting velocity resolution is \qty{8}{km\,s^{-1}}. 
The centroid velocity maps from KPNO employed in this work are based on the intensity-weighted mean velocities derived by \citet{Garcia-Diaz:2008a}. 
Their analysis incorporated additional east-west slit observations to improve the relative velocity calibration between adjacent slit positions.
The derived velocity fields were previously analyzed in \citet{arthur2016turbulence}.

Figure~\ref{fig:H_I-6563_maps} shows the two-dimensional maps of integrated intensity (left panel) and centroid velocity (right panel) for the Orion Nebula reconstructed using the MUSE data, both centered at the ionizing star of the region; the O7~V star $\theta^1$~Ori~C \citep{2006A&A...448..351S}. 
The field of view covers the brightest portion of the nebula commonly called the Huygens region.
Several well-known morphological features of the Orion Nebula can be identified in the left panel of the figure \citep{2009AJ....137..367O}. 

The region of highest surface brightness lies southwest of the Trapezium stars and corresponds to Orion-S, an active site of star formation and stellar outflows.
Within the Trapezium, $\theta^1$~Ori~C is the main source of ionizing radiation, accompanied by three cooler B-type stars.
Toward the southeast, the Orion Bar is visible, tracing the main ionization front of the nebula. 
To the northeast, the Dark Bay appears as a region of high extinction produced by neutral material in the Veil located on the observer's side of the nebula.

Several prominent kinematic features can be identified in the velocity map. 
The Herbig-Haro objects such as HH~201, HH~203, and HH~204 are traced by their strongly blue-shifted velocities \citep{Doi:2004a}. 
A further blue-shifted structure, the Big Arc South, is located south of Orion-S. 
At red-shifted velocities, the Red Bay appears northeast of this region, while the Red Fan is seen farther south \citep{Garcia-Diaz:2007a}.

\section{Methodology}\label{sec:met}

\subsection{Plane-of-sky velocity fluctuations}\label{sec:pos fluctuations}

Figure~\ref{fig:velocity_maps_shared_cbar_fixed} shows the VLT MUSE centroid velocity maps, \(\vcentroid\), for each emission line in our sample. 
These maps provide the basis for analyzing the velocity fluctuations projected onto the plane of the sky.
The centroid velocity maps used throughout this work are derived from the original observations after removing pixels associated with stellar sources identified from the continuum map. 
Using the \halpha\ emission line as a reference, we also discard pixels with normalized surface-brightness values below \(0.025\) and pixels associated with the high-velocity objects. 
The resulting spatial selection is then applied consistently to all remaining emission lines, ensuring that the turbulent analysis is performed over the same set of nebular regions. 
Each emission line is centered with respect it own mean centroid velocity value.
Pixels not included in the final velocity maps are shown in grey.
For comparison KPNO velocity maps are shown in Figure~\ref{fig:kpno vel maps}.

\begin{figure*}
 \centering
 \includegraphics[width=6in]{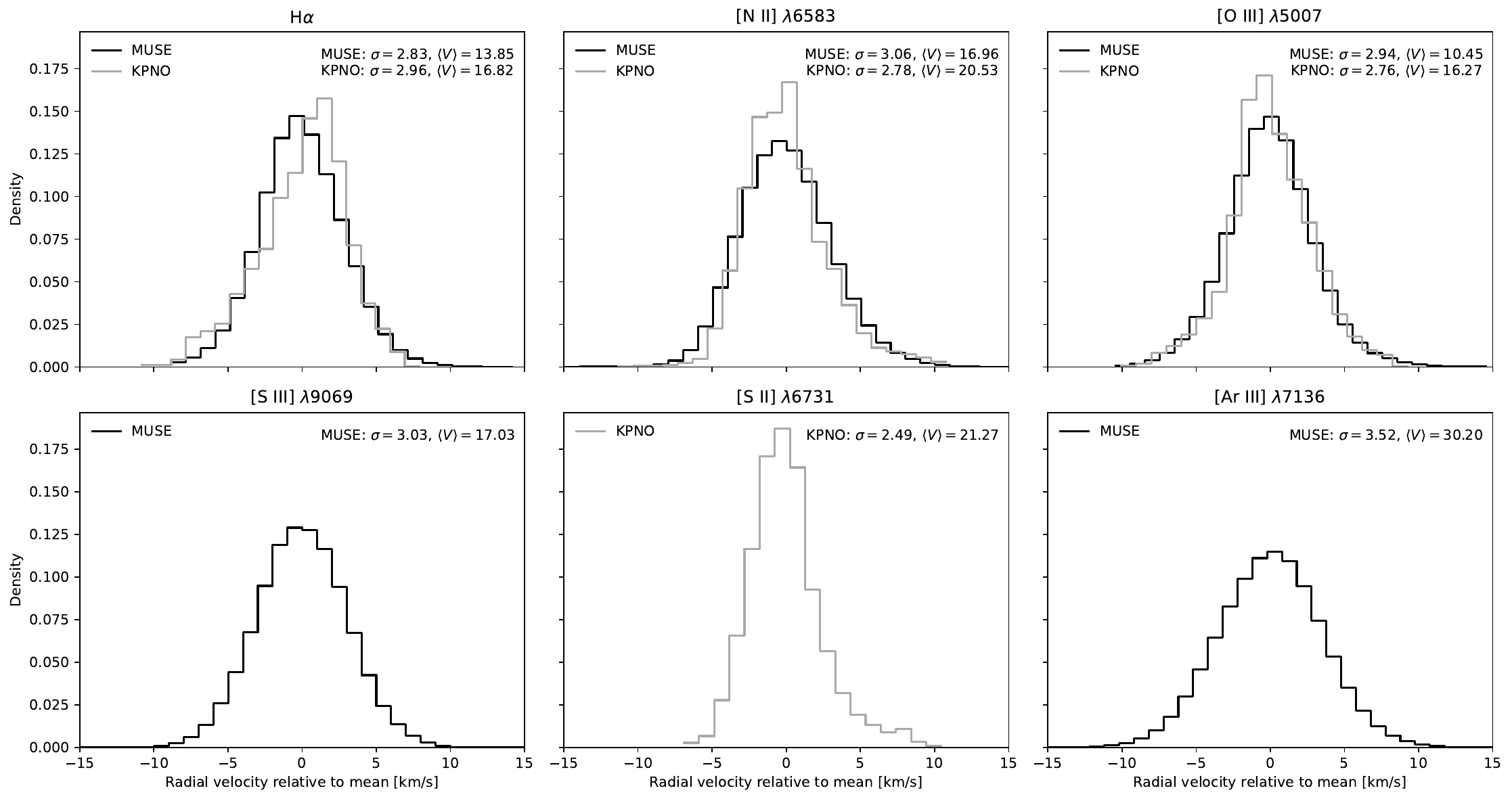}\par
 \caption{
   Probability density functions of the plane-of-sky centroid velocities for the different emission lines presented in Fig~\ref{fig:velocity_maps_shared_cbar_fixed}.
   The RMS width \(\sigma\pos\) of each distribution is shown.
   The bin width is \qty{1}{km.s^{-1}}.
 }
\label{fig:pdfs}
\end{figure*}

The overall amplitude of the plane-of-sky velocity fluctuations can be characterized by a dispersion, \(\sigma\pos\), defined as:
\begin{equation}
  \label{eq:sig-pos}
  \sigma^2\pos =
  \left\langle 
  \bigl[ \vcentroid (\xx_i) -\langle \vcentroid  \rangle  \bigr]^2
  \right \rangle ,
\end{equation}
where the average is performed over all observed points \(i\)
in a given map.
Note that \(\sigma\pos\) is also the RMS width of
the probability density function (PDF) of \(\vcentroid\).
The PDFs for all emission lines analyzed in this work are shown in Figure~\ref{fig:pdfs}.

The velocity PDFs of MUSE and KPNO data are shown after subtracting the mean velocity from each map, centering all distributions at \qty{0}{km\,s^{-1}}. 
They represent the amplitude distribution of the plane-of-sky velocity fluctuations and exhibit approximately Gaussian profiles for all emission lines. 
This behavior is expected after removing high-velocity features associated with Herbig-Haro objects and proplyds, together with low signal-to-noise pixels and regions contaminated by bright stars, resulting in more symmetric velocity distributions.
All PDFs span a similar velocity range, approximately from \qty{-10}{km\,s^{-1}} to \qty{10}{km\,s^{-1}} which display broadly similar kinematic behavior.
The \n\ emission line presents a small tail toward positive values which maps a southwest region in Figure~\ref{fig:velocity_maps_shared_cbar_fixed}. 
The broader distribution observed in the \ariii\ emission line is likely dominated by observational noise due to the intrinsic faintness of the line, whereas the narrower distribution of \halpha\ reflects its higher signal-to-noise ratio.

Although the PDFs characterize the overall amplitude of the velocity fluctuations through \(\sigma\pos\), they do not contain information about the spatial correlations of the velocity field. 
Therefore, a correlation analysis, such as the second-order structure function, is required to characterize the scale dependence of the turbulent fluctuations.
A summary of the velocity statistics is shown in Table~\ref{tab:velocity-statistics}.

\begin{table}
  \begin{center}\caption{ Summary of velocity statistics of VLT MUSE and KPNO data. }
\begin{tabular}{cCCC}\toprule
Data     & \text{Emission}    &  \langle V \rangle \pm \sigma           & N_\text{points}  \\
         &  \text{line}       &  [\si{km.s^{-1}}]     &  [-]              \\ 
\midrule
MUSE & $\halpha$ & 14 \pm 3  &     2187334          \\
     & $\n$      & 17 \pm 3 &     2183754            \\
     & $\oiii$   & 10 \pm 3 &    2187655         \\
     & $\siii$   & 17 \pm 3  & 2183754          \\
     & $\ariii$  & 30 \pm 4  &    2187355         \\
KPNO & $\halpha$ & 17 \pm 3  &    179958          \\
     & $\n$      & 20 \pm 3  &     179862            \\
     & $\oiii$   & 16 \pm 3  &     180156        \\
     & $\sii$    & 21 \pm 3   &   102342          \\
\bottomrule
\end{tabular}\label{tab:velocity-statistics}
\end{center}
\end{table}


\subsection{The binning procedure}\label{sec:binning procedure}

To mitigate the impact of instrumental noise while preserving the underlying turbulent signal, we apply a spatial binning procedure to the velocity maps using the \texttt{tetrablok} Python package\footnote{\url{https://github.com/will-henney/tetrabloks}}. 
This approach explicitly addresses the trade-off between noise reduction and spatial resolution in the recovery of the turbulent structure function \citep{2018MNRAS.479.3909G}.

The binning is performed by re-sampling the original data onto progressively coarser grids using binning factors of \(2^1 = 2\), \(2^2 = 4\), \(2^3 = 8\), and \(2^4 = 16\) applied in each spatial direction.
When passing from a finer grid to the next coarser grid,
the flux-weighted average velocity is calculated over all valid pixels in a \(2 \times 2\) block,
but only if the number of valid pixels is \(\ge N_{\text{min}}\), where \(N_{\text{min}}\) is a tunable parameter.
In this paper, we adopt \(N_{\text{min}}=1\), which means that missing data on the original grid
tends to get filled in on the coarser grids.

\begin{figure*}
 \centering
 \includegraphics[width=6.5in]{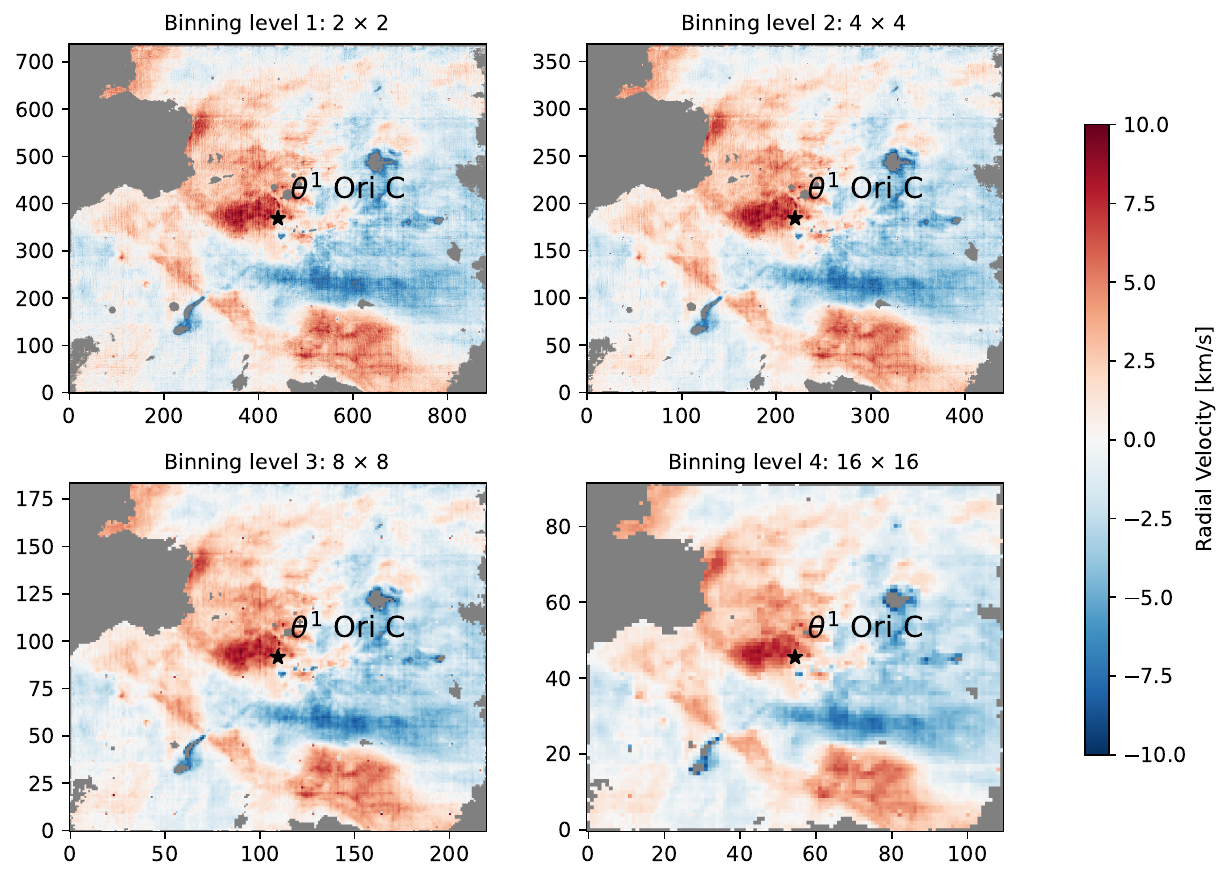}\par
 \caption{
 Two-dimensional binned maps of the centroid velocity of the Orion nebula for the \halpha\ emission line. The size of each map are \(737 \times 882\), \(368\times 440\), \(183 \times 219\)  and \(91 \times 109 \ \text{pixels}\) which correspond to the bin sizes of $2^1=2$, $2^2=4$, $2^3=8$ and $2^4=16$ respectively. The same procedure of binning is applied to all emission lines.  
 }
\label{fig:Ha_bins_comparison}
\end{figure*}

Figure~\ref{fig:Ha_bins_comparison} shows as an example of the binned velocity maps for the \halpha\ emission line.
The original map has a size of \(1 475 \times 1765\) original pixels
(see Fig.~\ref{fig:H_I-6563_maps}) which is reduced to a size of \(91 \times 109\)
coarse pixels
through a series of four iterations.
Each iteration and the resulting velocity map is shown in the different panels of Fig.~\ref{fig:Ha_bins_comparison} for the \halpha\ emission line.

\subsection{The second-order structure function}
\label{sec:second-order-struct}

After the binning and masking of the maps we apply the second-order structure function, \(B(r)\), which is a function of the scalar separation or lag, \(r\),
between two points on the plane of the sky:
\newcommand\Abs[1]{\vert #1\vert}
\begin{equation}\label{eq:Br}
  B(r) = \left\langle 
  \bigl[
  V_{c}(\xx_j) - V_{c}(\xx_i)
  \bigr]^{2} \right \rangle_{\Abs{\xx_j - \xx_i\!} \ \approx \ r} \ .
\end{equation}

The averaging is performed over all pairs of points
\((i, j)\)
whose scalar separation \(\Abs{\xx_j - \xx_i}\) is close to \(r\),
irrespective of the orientation of the separation vector.
In practice, we achieve this by binning the separations with a constant
logarithmic width of \qty{0.05}{dex}.

%
%

To interpret the observed structure function and extract the underlying turbulent parameters, we adopt a parametric model that accounts for both intrinsic turbulence and observational effects \citep{garciav23}. 
In this framework, the observed structure function is modeled by combining the intrinsic structure function with the effects of atmospheric seeing, \(S(r)\), and a constant noise contribution, \(B\noise\), yielding the corrected model:
\begin{equation}
  \tilde{B}\model(r) = B\model(r) \,  S(r) + B\noise
  \label{eq:sf-functional}
\end{equation}
where \(B\model(r)\) represents the intrinsic (noise- and seeing-free) structure function, which we model as:
\begin{equation}
  \label{eq:model-strucfunc-ideal}
  B\model(r) = 2\sigma^2\pos \left[
    1 - 2^{- \left( r/r_0 \right)^m} \ 
  \right] .
\end{equation}
Here, \(r_0\) is the correlation length, the characteristic length of the velocity fluctuations, and \(m\) is the is the power-law index, which characterize the slope of the turbulent cascade.
The model in equation~\eqref{eq:model-strucfunc-ideal} corresponds to the \textit{stable covariance model}, a standard model used in the field of geostatistics \citep{wackernagel2003multivariate}.

In equation~\eqref{eq:model-strucfunc-ideal}, the seeing \(S(r)\) is given by:
\begin{equation}\label{eq:ffs}
  S(r) =
  \left[ \frac{1}{1 + 1.25 s_0 / r_0} \right]
  \left[ \frac{1}{1 + (2.6 s_0 / r)^{1.5}} \right]
  ,
\end{equation}
where \(s_0\) is the RMS width of the seeing profile\footnote{%
  Note that the full-width half maximum (FWHM) seeing width is
  \(2 (2 \ln 2)^{1/2} s_0 \approx 2.35 s_0\).
}.

The observed structure function, \(B\obs(r)\), is computed using equation~\eqref{eq:Br}. 
The uncertainties in \(B\obs(r)\) are assumed to be dominated by systematic effects rather than random noise. Consequently, we assign an uncertainty to each bin equal to a constant fraction (1–3\%) of the observed value.

Model fitting is performed using a non-linear weighted least-squares approach based on the Levenberg–Marquardt algorithm \citep{More:1978a}, as implemented in the \texttt{lmfit} Python library \citep{newville_matthew_2014_11813}. 
The posterior distributions of the model parameters are estimated using Markov Chain Monte Carlo (MCMC) ensemble sampling \citep{2010CAMCS...5...65G}, implemented in the \texttt{emcee} Python library \citep{2013PASP..125..306F}.  
Table~\ref{tab:parameter-ranges} lists the uniform prior distributions assumed for each parameter.

The model in equation~\eqref{eq:sf-functional} assumes a statistically homogeneous turbulent velocity field, for which the autocorrelation function remains positive and decreases monotonically with separation.
At the largest spatial scales this assumption may break down owing to large-scale inhomogeneities in the nebula.
Rather than adopting a single fitting range, we therefore repeat the analysis using maximum separations of \(0.4L\), \(0.5L\), and \(0.6L\), where \(L\) is the size of the observed field.
The final parameter estimates are taken as the average of the three fits, while the quoted uncertainties encompass both the posterior uncertainties from each fit and the variation between the three fitting ranges.

To assess the quality of the fits, we consider the turbulent parameters to be reliably recovered when a power-law regime is identified and the derived parameters are well constrained. For this, we require statistical consistency between the model and the data, quantified by a reduced chi-square value \(\chi^2 \lesssim 1\). 
We also verify that the posterior distributions of the fitted parameters are well-behaved within the 95\% confidence interval, without significant asymmetries or multimodalities.

\begin{table}
  \centering
  \caption{Bounds of allowed values for parameters in model fits. The quantity \(L\) denotes the size of the observational box.}
  \label{tab:parameter-ranges}
  \newlength\partabwidth
  \setlength\partabwidth{0.8\linewidth}
  \begin{tabular*}{\partabwidth}{
    l @{\extracolsep{\fill}}
    r 
    r
    }
    \toprule
    Parameter & Lower & Upper\\
    \midrule
    \(\sigma^2\) & \(0.25\, \max [B\obs]\)& \(2\, \max [B\obs]\)\\
    \(r_0\) & \(0.01\, L\) & \(2\, L\)\\
    \(m\) & \(0.5\) & \(2.0\) \\
    \(s_0\) & \SI{0.1}{arcsec}& \SI{1.5}{arcsec}\\
    \(B\noise\) & \(0\) & \(3\, \min [B\obs]\) \\
    \bottomrule
    \multicolumn{3}{@{}p{\partabwidth}@{}}{
    Note: \(\max[B\obs]\) and \(\min[B\obs]\) are over all bins in the observed structure function with \(r < L/2\).
    }
  \end{tabular*}
\end{table}

\section{Results}\label{sec:results}


\begin{figure*}
 \centering
 \includegraphics[width=6.5in]{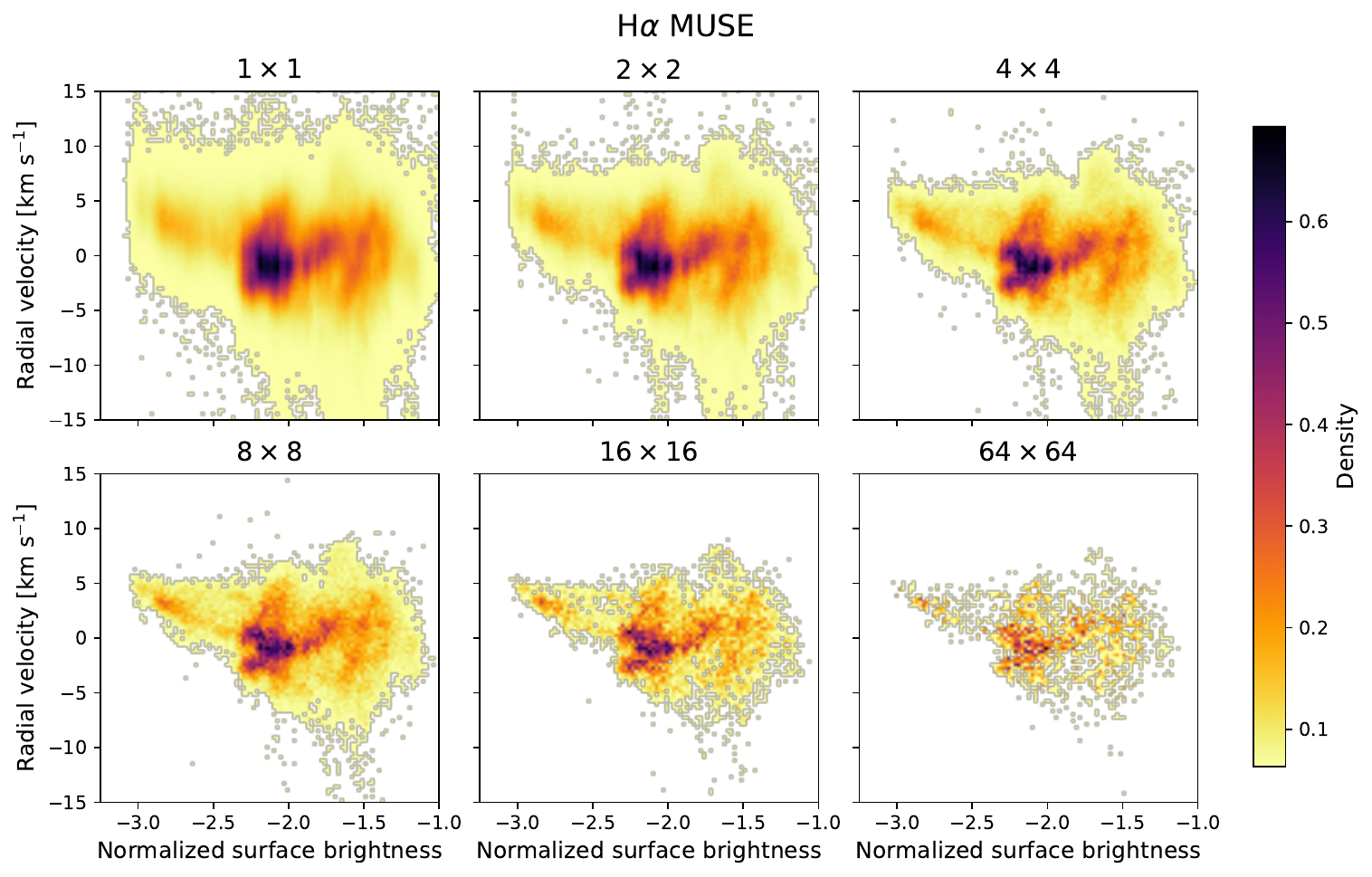}
 \caption{
   Joint distribution histograms of surface brightness and relative radial velocity centroid of \halpha{}
   from the full MUSE field for different binning levels.
   The darkness of the colored images is proportional to the number of pixels with
   each combination of brightness and velocity.
   No masking was applied, so all pixels from each dataset were used to construct the diagrams.
   The surface brightness values are normalized to the median (multiplied by \num{100}) of each dataset.
 }
 \label{fig:intensity_velocity_ha}
\end{figure*}

\begin{figure*}
 \centering
 \includegraphics[width=6.5in]{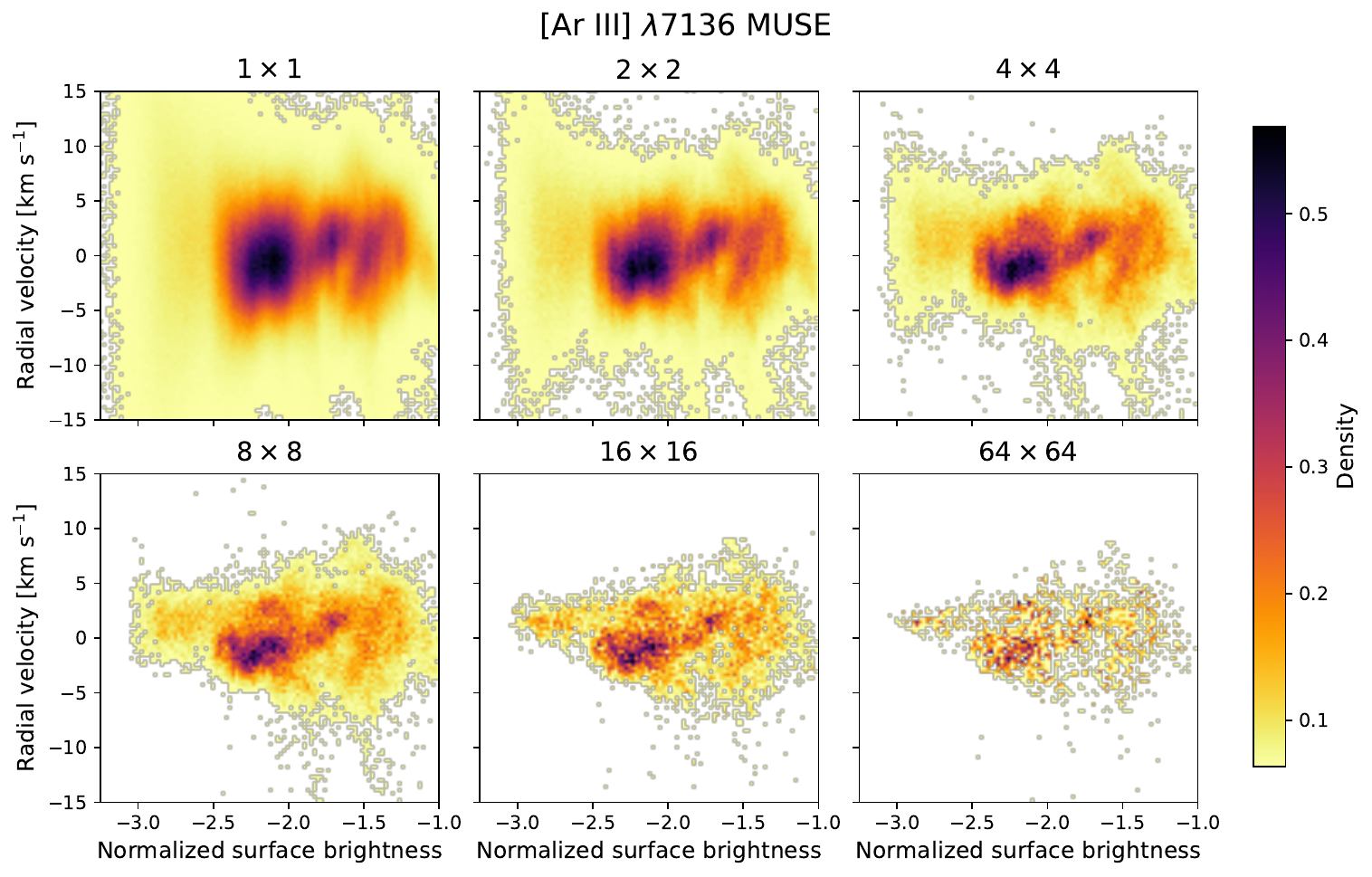}
 \caption{
   As Fig.~\ref{fig:intensity_velocity_ha} but for the \ariii{} line.
 }
 \label{fig:intensity_velocity_ar}
\end{figure*}

\subsection{Surface brightness vs radial velocity diagrams}\label{sec:sb vs velocity diagrams}

Figure~\ref{fig:intensity_velocity_ha} shows the joint distribution histograms of surface brightness and centroid velocity (relative to mean)
for the \halpha{} line from the full MUSE field shown in Fig.~\ref{fig:H_I-6563_maps},
with each panel representing a different binning level, as labelled.
These diagrams provide a direct view of the relation between line brightness and centroid velocity for each emission line, allowing us to compare the velocity distribution of bright and faint gas across datasets.
The unbinned data (upper left panel) appears considerably blurred along the vertical (velocity) axis,
due to the per-pixel noise in the velocity map.
As the binning is increased to \(2\times 2\) and \(4 \times 4\) (upper row of panels),
this blurring is considerably decreased and the underlying structure resolves into sharper focus.
This is most notable at low surface brightness (left-hand side of each panel),
where the distribution in velocity becomes very narrow at higher binning levels.
In contrast, the broad velocity distribution at high surface brightness,
including an extended wing towards negative velocities, is relatively unaffected by the binning,
indicating that it is physically real, rather than due to noise.
As the binning is increased further through \(8 \times 8\) and \(16 \times 16\) to \(32 \times 32\),
the sharpness of the structures stops improving and instead the decreasing number of map pixels
means that the histograms themselves become increasingly noisy due to counting-statistics fluctuations.

Figure~\ref{fig:intensity_velocity_ar} shows the same joint histograms but for the \ariii{} line.
This line is roughly 20 times fainter than \halpha{} (Fig.~\ref{fig:ratios}),
so the noise level in the velocity map is much higher, as can be seen directly in Fig.~\ref{fig:velocity_maps_shared_cbar_fixed}.
As a result, the blurring along the velocity axis of the unbinned data is more extreme,
although it is reduced as the binning level increases,
with the high surface brightness side becoming sharp by the \(4 \times 4\) binning level,
while the low surface brightness side remains noise-broadened up to the \(16 \times 16\) binning level.

\begin{figure*}
 \centering
 \includegraphics[width=6.5in]{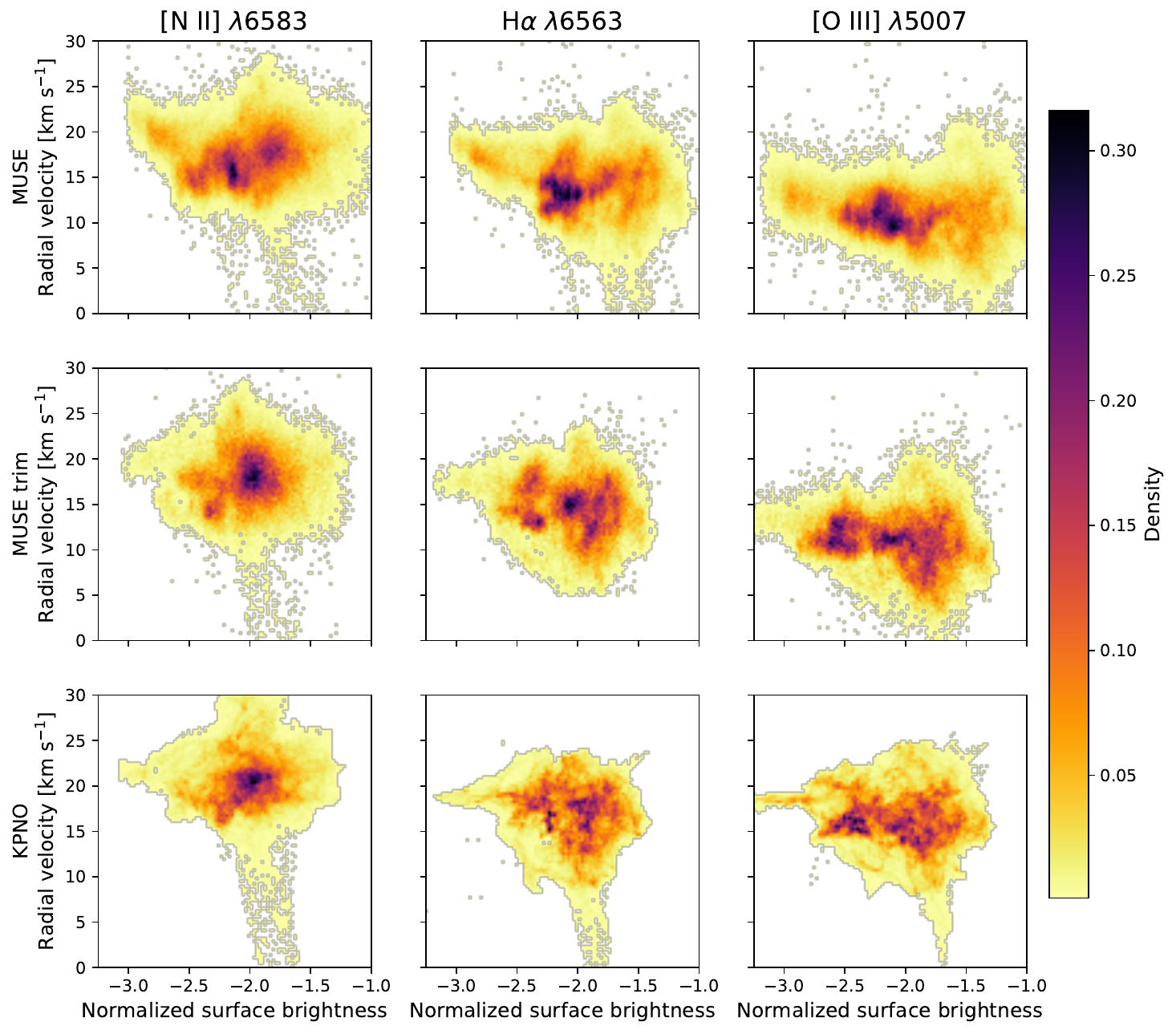}\par
 \caption{
   As Fig.~\ref{fig:intensity_velocity_ha} but comparing different emission lines (columns)
   and observational datasets (rows).
   Joint distribution histograms of surface brightness and radial velocity centroid
   (heliocentric scale)
   for  \n{} (first column), \halpha{} (second column), and \oiii{} (third column). 
   The first row shows the full VLT MUSE field of view with a \(4\times4\) binning. 
   The second row shows the VLT MUSE data restricted to the KPNO field of view, also with a \(4\times4\) binning. 
   The third row shows the KPNO data with no binning. 
 }
\label{fig:intensity_velocity}
\end{figure*}

Figure~\ref{fig:intensity_velocity} compares the joint distributions between the
VLT MUSE and KPNO datasets, showing the emission lines in common between the two:
\n{}, \halpha{}, \oiii{}.
The spatial binning level is held constant at \(4 \times 4\) for the MUSE data,
whereas no binning is used for the KPNO data, which are already at a lower spatial resolution.
The first row shows the full VLT MUSE field of view, while the second row shows
the VLT MUSE data restricted to the KPNO field of view, which we denote MUSE-trim, and the third row shows the KPNO observations.
Visual comparison of the bottom two rows shows a remarkable consistency between the MUSE and KPNO
results for each emission line, when restricted to the same field of view.
Fine details in the joint distributions can be matched between the two datasets, but there are two main differences,
which indicate the limitations of the data.
First, there is a systematic offset between the MUSE and KPNO velocities of \qtyrange{3}{6}{km.s^{-1}},
which is not unexpected given the finding by \citet{2015A&A...582A.114W} of a similar offset between MUSE velocities
and values from the literature \citep{Baldwin:2000a}.
The precision of the KPNO absolute velocities is estimated to be better than this, at \(\approx \qty{2}{km.s^{-1}}\) \citep{Garcia-Diaz:2008a}.
Second, the KPNO distributions appear to be slightly compressed along the brightness axis,
compared with the MUSE distributions.
This is probably due to the fact that the MUSE spectra are sky-subtracted,
while the KPNO spectra are not,
with the resultant difference in the brightness zero point manifesting
as stretching/compression on a log scale.
Since this paper concentrates on relative velocity fluctuations and employs the surface brightness
only for masking the velocity maps,
our results are not significantly affected by either of these issues.

A third difference is that the MUSE joint distribution histograms in Figure~\ref{fig:intensity_velocity}
appear more blurred than the KPNO ones, especially along the velocity axis,
which is a direct result of the larger uncertainty in the radial velocity measurements of the MUSE observations.
As shown above (Figs.~\ref{fig:intensity_velocity_ha}, \ref{fig:intensity_velocity_ar}),
this can be improved by increasing the binning level of the maps,
but only at the cost of worsening the counting-statistics fluctuations of the histograms
due to the reduced number of map pixels.

\subsection{Noise-resolution trade off}\label{sec:noise-resolution trade off}

In this section we show the turbulent parameters and their confidence intervals obtained by fitting equation~\eqref{eq:sf-functional} to the observed structure function, \(B\obs(r)\). 
We first present the results for the \halpha\ and \ariii\ emission lines, which correspond to the brightest and faintest lines in the MUSE sample, respectively. 
These two cases illustrate the effects of masking and binning on the structure function under different signal-to-noise conditions. 
This is followed by the same analysis applied to the remaining emission lines, with their respective results presented in the Appendix~\ref{apex:multiple_lines_results}. 

\begingroup
\setlength{\tabcolsep}{6pt} 
\renewcommand{\arraystretch}{1.5} 
\begin{table*}
\begin{center}
  \caption{
    Best-fit model parameters and credibility intervals for fits to observed structure functions in the Orion core for the VLT MUSE \halpha\ line observations.
  }

  
\begin{tabular}{l RRR @{\hspace{6\tabcolsep}} RR @{\hspace{6\tabcolsep}} RRR}
  \toprule
Binning 
& \sigma^2\pos
& r_0
& m
& B_{\text{noise}}
& s_0 (\text{FWHM})
& \text{SNR}
& s_0 / r_0
& \text{Bin size} / r_0  \\
level
& [\si{km^2.s^{-2}}]
& [\si{pc}]
& [-]
& [\si{km^2.s^{-2}}]
& [\si{arcsec}]
& [-]
& [-]
& [-] \\
\midrule
1$\times$1 &
8.94\PM{1.18}{0.65} & 0.069\PM{0.014}{0.007} & 1.13\PM{0.07}{0.08} & 2.614\PM{0.106}{0.067} & <0.92 & 2.61 & <0.0106 & 0.005 \\

2$\times$2 &
8.91\PM{0.68}{0.51} & 0.068\PM{0.007}{0.006} & 1.15\PM{0.05}{0.04} & 1.222\PM{0.081}{0.040} & <0.86 & 3.82 & <0.0102 & 0.011 \\

4$\times$4 &
8.88\PM{0.61}{0.50} & 0.068\PM{0.007}{0.005} & 1.16\PM{0.05}{0.04} & 0.549\PM{0.098}{0.051} & <1.23 & 5.69 & <0.0144 & 0.022 \\

8$\times$8 &
8.71\PM{0.53}{0.41} & 0.066\PM{0.005}{0.004} & 1.20\PM{0.05}{0.05} & 0.332\PM{0.160}{0.084} & <2.37 & 7.25 & <0.0287 & 0.046 \\

16$\times$16 &
8.65\PM{0.51}{0.44} & 0.066\PM{0.004}{0.004} & 1.25\PM{0.07}{0.06} & 0.227\PM{0.203}{0.198} & <3.45 & 8.74 & <0.0419 & 0.091 \\

\bottomrule

\end{tabular}\label{tab:results_MUSE_Ha}
\end{center}
\end{table*}
\endgroup

\begingroup
\setlength{\tabcolsep}{6pt} 
\renewcommand{\arraystretch}{1.5} 
\begin{table*}
\begin{center}
  \caption{
    Best-fit model parameters and credibility intervals for fits to observed structure functions in the Orion core for the VLT MUSE \ariii\ line observations.
  }

  
\begin{tabular}{l RRR @{\hspace{6\tabcolsep}} RR @{\hspace{6\tabcolsep}} RRR}
\toprule
Binning
& \sigma^2\pos
& r_0
& m
& B_{\text{noise}}
& s_0 (\text{FWHM})
& \text{SNR}
& s_0 / r_0
& \text{Bin size} / r_0  \\
level
& [\si{km^2.s^{-2}}]
& [\si{pc}]
& [-]
& [\si{km^2.s^{-2}}]
& [\si{arcsec}]
& [-]
& [-]
& [-] \\
\midrule
1$\times$1
& 8.45\PM{0.60}{0.37} & 0.067\PM{0.007}{0.004} & 1.11\PM{0.06}{0.07} & 10.806\PM{0.218}{0.076} & <2.65 & 1.25 & <0.0318 & 0.006 \\

2$\times$2
&8.94\PM{0.35}{0.35} & 0.062\PM{0.004}{0.004} & 1.11\PM{0.05}{0.03} & 4.120\PM{0.169}{0.042} & <1.37 & 2.08 & <0.0176 & 0.012 \\

4$\times$4
&9.36\PM{0.51}{0.47} & 0.061\PM{0.005}{0.005} & 1.06\PM{0.05}{0.04} & 1.569\PM{0.192}{0.098} & <1.52 & 3.45 & <0.0200 & 0.025 \\

8$\times$8
&9.76\PM{0.65}{0.58} & 0.058\PM{0.006}{0.005} & 1.02\PM{0.07}{0.05} & 1.006\PM{0.378}{0.251} & <2.76 & 4.40 & <0.0381 & 0.052 \\

16$\times$16
&9.87\PM{0.65}{0.59} & 0.054\PM{0.005}{0.004} & 1.02\PM{0.10}{0.07} & 0.397\PM{0.821}{0.277} & <3.46 & 7.05 & <0.0515 & 0.112 \\

\bottomrule

\end{tabular}\label{tab:results_MUSE_Ar}
\end{center}
\end{table*}
\endgroup

\begin{figure*}
 \includegraphics[width=0.48\textwidth]{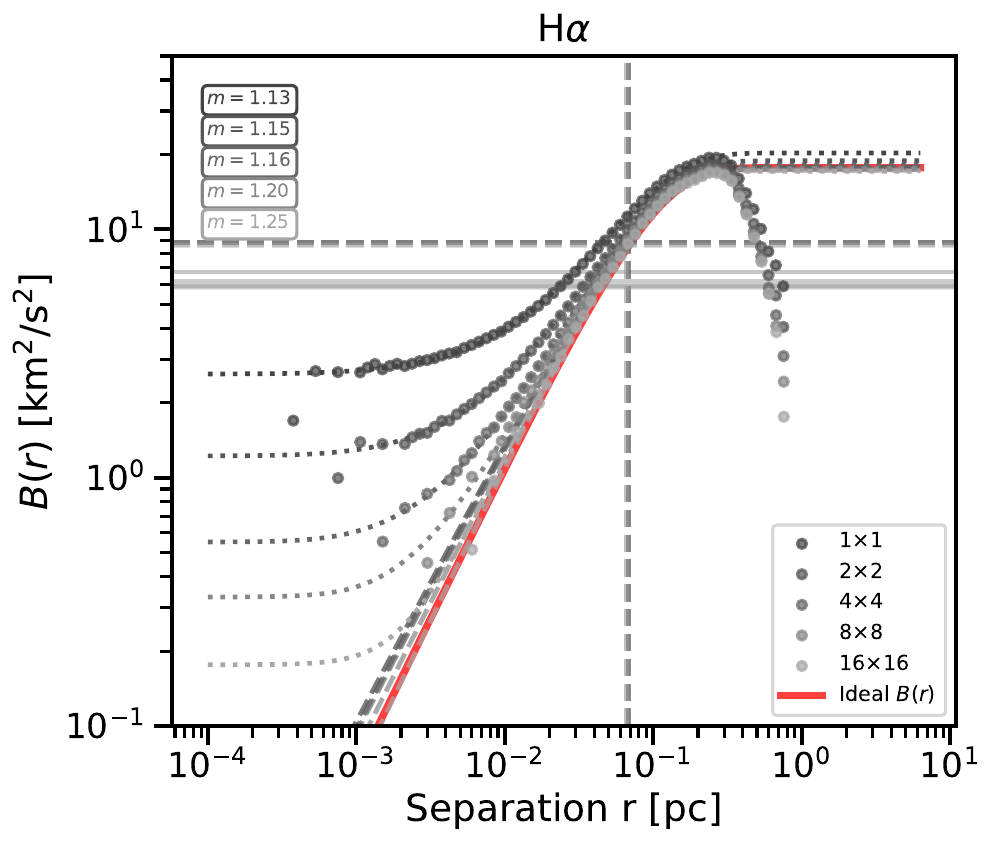}
 \hfill
 \includegraphics[width=0.48\textwidth]{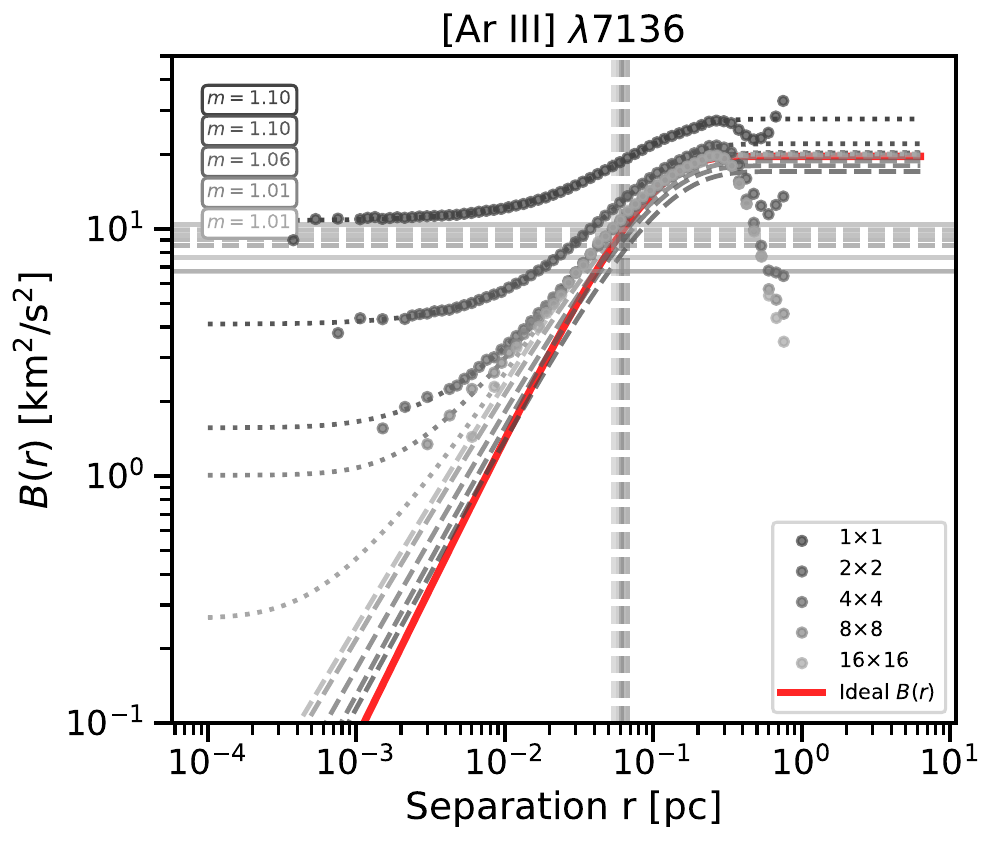}
 \caption{
   Second order structure functions of the velocity centroid derived from MUSE observations
   of the Orion Nebula.
   (a)~\ha. (b)~\ariii.
   The solid markers show the observational structure function computed using equation~\eqref{eq:Br}, for the non-binned map ($1\times1$) and binned maps ($2\times2$, $4\times4$, $8\times8$ and $16\times16$). 
   The dotted line corresponds to the heuristic model described in equation~\eqref{eq:model-strucfunc-ideal}, while the dashed lines represent the underlying model given in equation~\eqref{eq:sf-functional}. 
   The vertical dashed line shows the value of the correlation length \(r_0\) and the horizontal dashed lines shows the value of the \(\sigma^2\pos\) recover from the fit.
   The solid horizontal lines how the value of \(\sigma^2\) obtained form the observations and binning.
   The results in the figures correspond to fits performed up to a maximum separation of \(0.5L\).
   The plot illustrates how the binning procedure reduces the impact of instrumental noise at small separations, as indicated by the dotted curve. Notably, the non-binned map still allows a reliable recovery of the turbulent parameters, demonstrating that the methodology remains effective even under low signal-to-noise conditions.
   The thick red line corresponds to the ideal underlying  \halpha\ structure function, equation \eqref{eq:model-strucfunc-ideal}, derived from the KPNO observations (rescaled to the same \(r_0\) and \(\sigma^2\) as the MUSE observations), which serve as the reference dataset.
 }
 \label{fig:Ha_Br_comparison}
\end{figure*}

\subsubsection{Representative case study of the \halpha\ line}
\label{sec:Representative case study}

Empirical and model structure functions for the \halpha{} velocity fluctuations are shown in Figure~\ref{fig:Ha_Br_comparison}a.
Solid markers of different colors indicate the results of applying equation~\eqref{eq:Br} to the non-binned \halpha\ map shown in Figure~\ref{fig:velocity_maps_shared_cbar_fixed}, and the corresponding binned maps in Figure~\ref{fig:Ha_bins_comparison}.
The dotted curves correspond to the best-fit heuristic model described in equation~\eqref{eq:sf-functional}, while the dashed curves represent the underlying model given in equation~\eqref{eq:model-strucfunc-ideal}
(that is, ignoring the effects of seeing and noise).  
For each binning level the size of the pixel in parsecs ranges from \(\binsize = 0.0004\) (unbinned)
to \(0.0064\) (binning \(16 \times 16\)).
For comparison, the red solid line in the figure shows the ideal underlying model derived from
the higher spectral resolution KPNO observations (Fig.~\ref{fig:fit_results_04}a),
rescaled to the same \(r_0\) and \(\sigma^2\) as the MUSE observations to account for the different field of view. 

For the \halpha\ emission we were able to fit the model to all the observed structure functions. Table~\ref{tab:results_MUSE_Ha} lists the best-fit values for the model parameters with 95\% confidence intervals.
In addition to the turbulence parameter (columns 2--4)
and nuisance parameters (columns 5--6),
the table shows
the ratios \(s_0/r_0\) and \(\binsize{} / r_0\), which are used for a sanity check of the results.
We also estimate the signal to noise ratio, SNR, of the turbulent fluctuations as:
\begin{equation}\label{eq:SNR}
  \text{SNR} = \frac{\sigma\pos}{\sigma_\text{noise}} = \left( \frac{2 \sigma\pos^2}{B_\text{noise}} \right)^{0.5}.
\end{equation}
The variation of the model parameters with binning level is also shown graphically in Figure~\ref{fig:parameters vs binning}.

For the non-binned \halpha\ map, a power-law trend is present, but the structure at small scales is significantly affected by instrumental noise, with a derived value of \qty{2.61}{km^{2}.s^{-2}}.  
The structure functions derived from the binned maps are also shown in Figure~\ref{fig:Ha_Br_comparison}.  
As the bin size increases, from \(2\times2\) to \(16\times16\), the small-scale structure of the velocity field becomes progressively clearer, as evidenced by the emergence of a well-defined power-law behavior and a reduction of the instrumental noise to a value of \qty{0.39}{km^{2}.s^{-2}}.  
For per-pixel uncorrelated noise, the variance of the rebinned map is expected to decrease as \(N_{\mathrm{pix}}^{-1}\), where \(N_{\mathrm{pix}}\) is the number of pixels within the bin. 
However, the observed decline in \(B_\text{noise}\) is less steep than this, indicating that the ``noise'' is partially spatially correlated and dominated by systematic instrumental effects rather than Poisson statistics.

As demonstrated in section~\ref{sec:sb vs velocity diagrams}, at a binning level of \(8 \times 8\) and above,
the binning begins to remove physically relevant small-scale structure.
In the model fits, this is manifested as an increase in the \(s_0\) parameter
(Fig.~\ref{fig:parameters vs binning}),
which was designed to model atmospheric seeing but in this case is responding to the effective smoothing induced by the binning.
Above the same binning level, the derived noise also stops decreasing, indicating
that \(4 \times 4\) is the optimum binning level for this dataset,
which corresponds to \(\binsize{} < 0.05 r_0\).

Although the noise reduction induced by the binning causes a significant cosmetic improvement
in the empirical structure function (symbols in Fig.~\ref{fig:Ha_Br_comparison}a),
it is important to note that the model recovers very similar turbulent parameters
at all binning levels (dashed lines in  Fig.~\ref{fig:Ha_Br_comparison}a).
As can be seen from Table~\ref{tab:results_MUSE_Ha} and Figure~\ref{fig:parameters vs binning},
the physical parameters \(\sigma^2\), \(r_0\), and \(m\) remain constant for binning levels
up to \(4 \times 4\).
For larger binning levels, we start to see slight reductions in \(\sigma^2\) and \(r_0\),
together with a steady increase in \(m\), which are all symptoms of the suppression of small-scale fluctuations due to the binning.

This demonstrates that our methodology, based on fitting a heuristic model to the observational data, is sufficient to recover the turbulent parameters of the velocity field. 
In practice, this implies that either the non-binned maps can be used, at the expense of increased computational cost, or binned maps can be employed within the constraint imposed by the ratio \(\binsize{} / r_0 < 0.05\), without compromising the recovery of the turbulent cascade. 
This result extends the applicability of the method to datasets with low signal-to-noise ratios or limited spatial resolution.

At large spatial scales, the \halpha\ structure function shown in Figure~\ref{fig:Ha_Br_comparison} exhibits a downward hook, which reflects a reduction in the amplitude of the velocity fluctuations towards the edges of the observational field, see discussion in \citet{garciav23}.

\begin{figure*}
  \centering
  \includegraphics[width=0.8\linewidth]{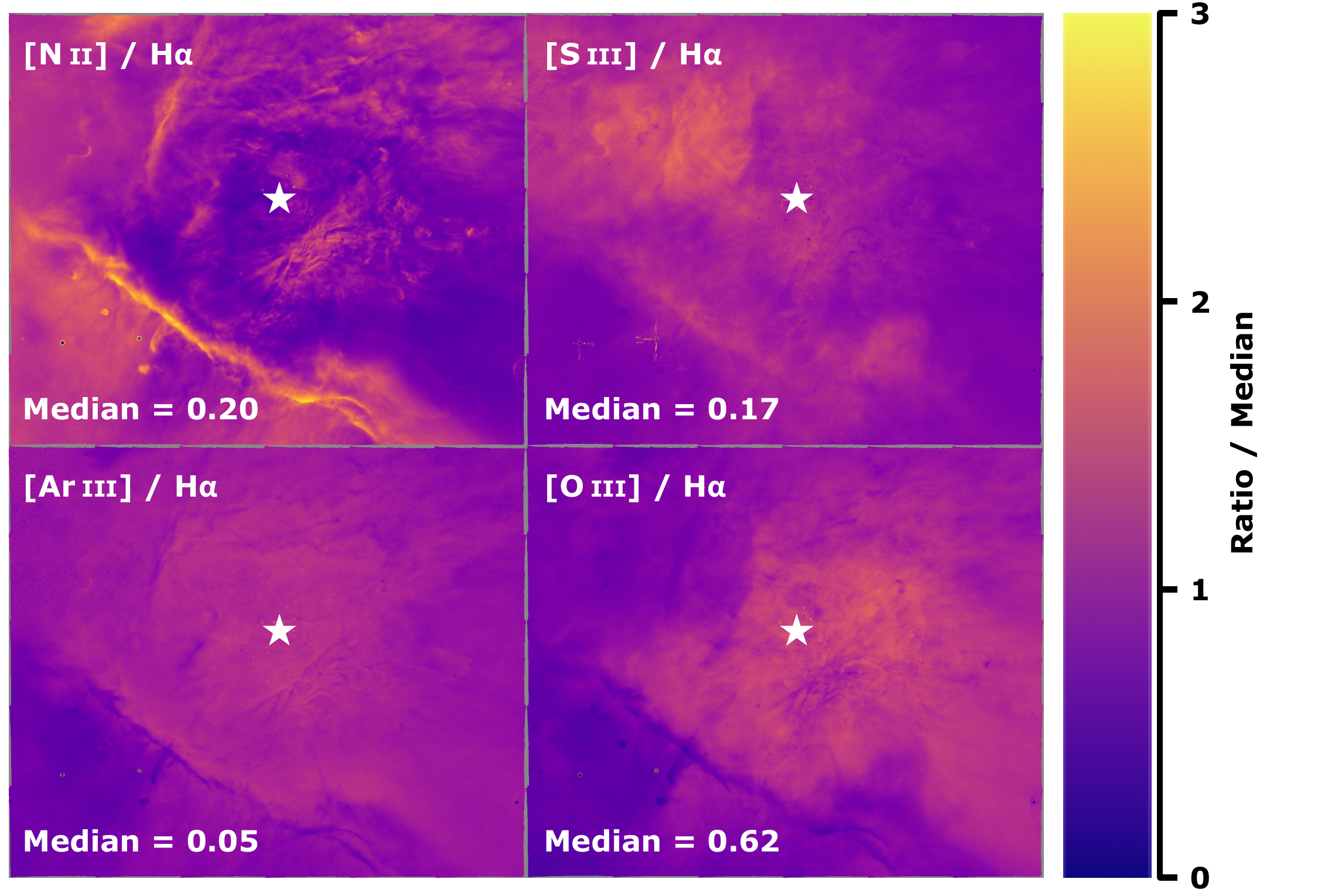}
  \caption{Surface brightness ratio maps of all other emission lines with respect to \halpha{}.
    The color scale for each map ranges from zero to three times the median ratio for each line,
    as indicated. The \(6' \times 5'\) field of view of each panel is the same as in Fig.~\ref{fig:H_I-6563_maps},
    with the position of $\theta^1$\,Ori~C marked by a star.
  }
  \label{fig:ratios}
\end{figure*}

\subsubsection{Representative case study of the \ariii\ emission line}\label{sec:Representative case study Ar}

The \ariii{} velocity map (Fig.~\ref{fig:velocity_maps_shared_cbar_fixed})
appears very similar to \halpha{},
except for the increased noise,
as discussed in section~\ref{sec:sb vs velocity diagrams}.
The \(\ariii / \halpha\) ratio shows very little variation over the face of the nebula
(lower-left panel of Fig.~\ref{fig:ratios}),
falling from \(0.06 \pm 0.01\) in the central regions to \(0.04 \pm 0.01\) in the outskirts.
This line therefore provides a useful natural experiment for testing the recovery
of turbulent parameters in lower signal-to-noise conditions.

\begin{figure*}
 \centering
 \includegraphics[width=6in]{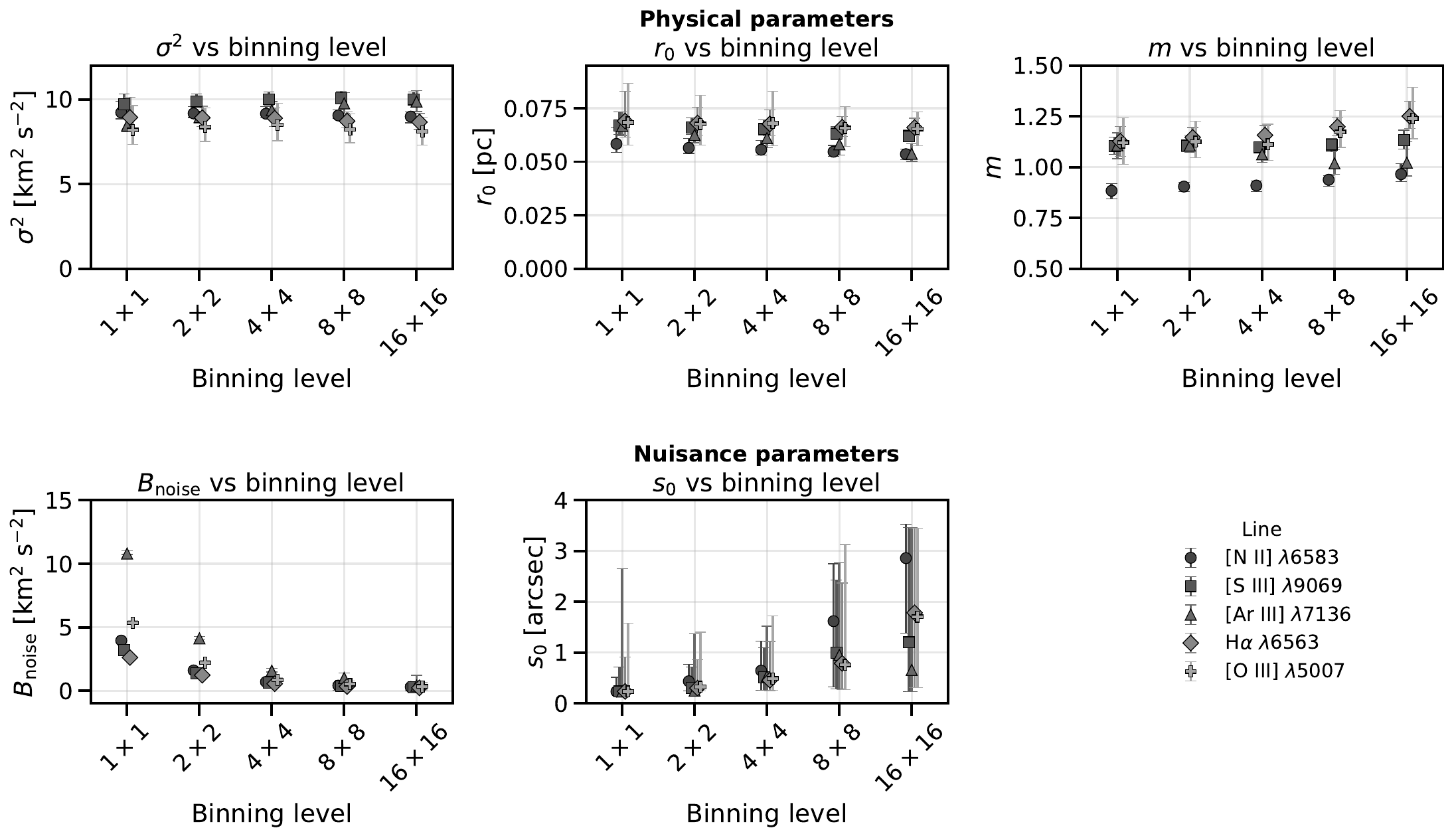}\par
 \caption{
Turbulent parameter versus binning level for all emission lines analyzed in this work and all binning levels for the VLT MUSE data.
 }
\label{fig:parameters vs binning}
\end{figure*}

Results for the empirically derived structure function and model fits
are shown in Fig.~\ref{fig:Ha_Br_comparison}b, with best-fit parameters listed in Table~\ref{tab:results_MUSE_Ar}.
As was apparent in the brightness-velocity histograms (Fig.~\ref{fig:intensity_velocity_ar}),
the unbinned velocity map is significantly affected by noise,
giving the underlying velocity fluctuations a SNR of order unity (Table~\ref{tab:results_MUSE_Ar}).
As a result, the empirical structure function is an overestimate of the true fluctuations at all scales.
Increased binning causes a rapid convergence of the empirical structure function
towards the ideal curve for separations \( > \qty{0.1}{pc}\),
but a much slower convergence at smaller scales.
The model fit parameters,
\(\sigma^2\), \(r_0\), and \(m\),
show larger variations than with \halpha{} (Fig.~\ref{fig:parameters vs binning}),
but these are still only of order 10\%.

\begin{figure*}
 \centering
 \includegraphics[width=6in]{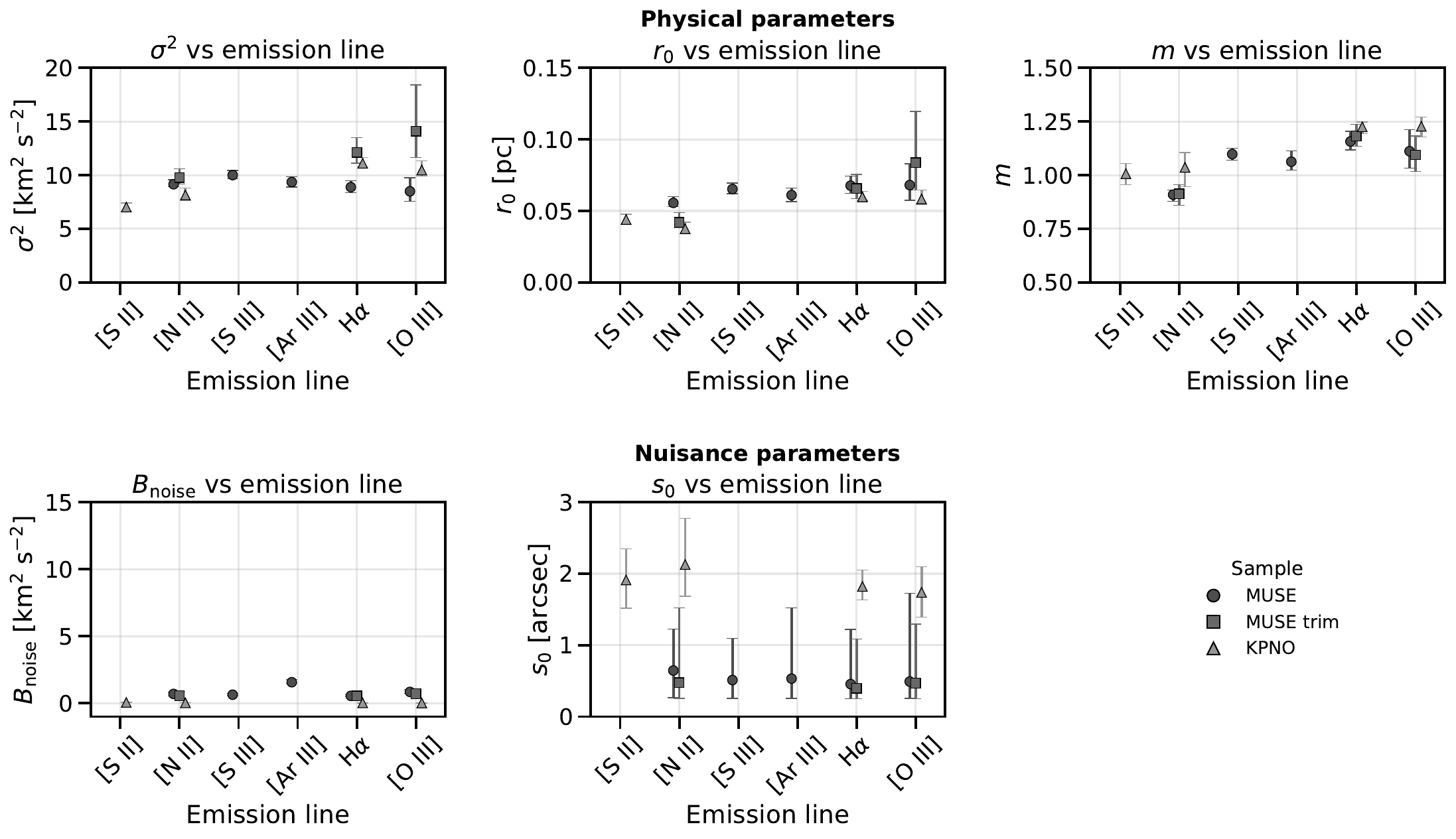}\par
 \caption{
 Turbulent parameter versus emission line for different samples analyzed in this work: MUSE (circles), MUSE using KPNO FoV (squares) and KPNO results (triangles). 
 The results presented here for the VLT MUSE and MUSE-trim data set correspond to binning levels of $4\times4$ for all emission lines.
 }
\label{fig:parameters_comparison}
\end{figure*}

\subsubsection{Extension to additional emission Lines}\label{sec:other emission lines}

The same analysis as for the \halpha\ and \ariii\ emission lines is extended to the remaining lines in the MUSE sample, with detailed results presented in
Figures~\ref{fig:fit_results_02}-\ref{fig:fit_results_03}
and Tables~\ref{tab:results_MUSE_O}-\ref{tab:results_MUSE_S} of
Appendix~\ref{apex:multiple_lines_results}.
The fitted parameters are summarized in Fig.~\ref{fig:parameters vs binning},
which shows the variation with binning level.
All these emission lines exhibit behavior consistent with the previous cases. 
The small-scale structure is progressively recovered as the binning level increases, accompanied by a systematic reduction in the instrumental noise and a corresponding increase in the effective seeing. 
From Figure~\ref{fig:parameters vs binning}, the noise decreases from typical values of \qty{\sim 5}{km^{2}.s^{-2}} to \qty{\sim 0.6}{km^{2}.s^{-2}} with increasing bin size (also see Tables~\ref{tab:results_MUSE_O}--\ref{tab:results_MUSE_S}). 
At the same time, the \(s_0\) parameter increases from values below \qty{1}{arcsec} for binning factors of \num{2} and \num{4}, to values above \qty{\sim 3}{arcsec} for bins \num{8} and \num{16}. 

Fig.~\ref{fig:parameters_comparison}
show the variation in fitted parameters between emission lines,
arranged in order of ionization potential of the parent ion,
including comparison with the KPNO results (Appendix~\ref{apex:kpno_structure_function}).
The three parameters \(\sigma^2\), \(r_0\), and \(m\) all show a weak increasing trend with ionization.


\section{Discussion}\label{sec:discussion}

By comparing velocity mapping of the Orion Nebula obtained from different observations with highly orthogonal instrumental limitations,
we can arrive at a more robust and complete characterization of the true velocity fluctuations in the nebula.
Note that in this paper, we are concerned solely with the statistics of the line centroid velocities \(\vcentroid\),
which are the result of a line-of-sight integration of the three-dimensional velocity field,
weighted by the emissivity of the particular line and potentially subject to radiative transfer effects,
such as absorption and scattering by dust \citep{1998ApJ...503..760H}.
The relation of the spatial statistics of the two-dimensional fluctuations in \(\vcentroid\)
to those of the three-dimensional fluctuations in velocity and emissivity is a complex topic,
which we defer to later papers,
but see discussion in \citet{garciav23}.

\subsection{Comparison between MUSE and KPNO results}

As presented in detail in section~\ref{sec:results},
the principal observational deficiency of the MUSE integral field spectroscopy is the
statistical uncertainty in the individual \(\vcentroid\) measurements, which we call ``noise''
and which ranges from \(\pm \qty{1}{km.s^{-1}}\) for the brighter lines
to  \(\pm \qty{3}{km.s^{-1}}\) for the fainter lines.
This has the effect of biasing upwards the empirical estimator of the structure function from its true value,
which is most noticeable at small separations, \(r\),
where it manifests as a flattening of the underlying power-law decline as \(r \to 0\).
We explore two different ways of compensating for the effects of noise: by modeling and by binning, neither of which are perfect.
The modeling (see eq.~\eqref{eq:sf-functional}) assumes that the statistical uncertainty in each pixel is independent and constant across the field,
neither of which are necessarily true.
The binning has the side-effect of suppressing real physical variations that might exist at small scales
and can also potentially introduce aliasing effects (cf.~App.~D of \citealp{Clerc:2019a}).
Despite these limitations, both approaches are successful in extracting a structure function power-law slope over the range \(r = \qtyrange{0.01}{0.1}{pc}\) of \(m \approx 1.1 \pm 0.1\)
for \halpha{} (see Fig.~\ref{fig:Ha_Br_comparison}),
which is consistent with the value previously obtained from KPNO observations
\citep{arthur2016turbulence, garciav23}.

In contrast, the principal observational deficiency of the KPNO longslit echelle spectroscopy
is the poor angular resolution due to the seeing and to the spatial interpolation between individual slits \citep{Garcia-Diaz:2008a}.
At small separations, this has the opposite effect from that of noise
and causes a steepening of the underlying power-law decline as \(r \to 0\).
This can be seen in Figs.~\ref{fig:fit_results_04} and \ref{fig:fit_results_05}
as a curving away of the empirical structure function below the power-law trend
for \(r \lesssim \qty{0.01}{pc}\).
The statistical uncertainty in individual \(\vcentroid\) measurement from the KPNO spectra
is of order \(\pm \qty{0.1}{km.s^{-1}}\),
which is much smaller than from MUSE and means that noise has a negligible effect on the empirical structure function.
We compensate for the effects of seeing by means of an empirical model (see eq.~\eqref{eq:ffs})
that has been calibrated on synthetic datasets (App.~A2 of \citealp{garciav23}),
although the lack of a physical justification for the functional form is a potential weakness
that we will address in the future.

The derived correlation length of \(r_0 \approx \qty{0.065}{pc}\), or \qty{36}{arcsec},
is also consistent between the MUSE and KPNO results,
but there is a difference of 20\% between the velocity variances:
\(\sigma^2 \approx \qty{9}{km^2.s^{-2}}\) for MUSE versus
\(\sigma^2 \approx \qty{11}{km^2.s^{-2}}\) for KPNO.
However, this is primarily due to the difference in field of view for the two datasets:
the height of the two fields (\qty{5}{arcmin} in N-S direction) is roughly the same,
whereas the MUSE field is roughly twice as wide (\qty{6}{arcmin} E-W direction) and includes regions in the outskirts of the nebula where the velocity fluctuations are slightly smaller.
To test this, we analyzed the subset of the MUSE maps that overlap with the KPNO fields (MUSE-trim),
finding a much closer agreement of \(\sigma^2\).
Full results of this analysis are listed in Table~\ref{tab:results_MUSE} and included in Figure~\ref{fig:parameters_comparison}.

\subsection{Comparison between emission lines}
\label{sec:comp-betw-emiss}

It is apparent from Fig.~\ref{fig:parameters_comparison} that there is only a modest variation
(of order 10\% or less) in the turbulent parameters between different emission lines.
On average, there is a general tendency for all the parameters,
\(\sigma^2\), \(r_0\), and \(m\),
to increase slightly with the degree of ionization,
but this is more pronounced for the smaller field of view of the KPNO observations.

\subsection{Behavior at large separations}
\label{sec:behavior-at-large}

Our model structure function saturates to a constant value for separations larger than about
twice the correlation length \(r_0\).
However, the empirical structure functions do not follow this behavior.
For the majority of emission lines observed by MUSE,
they reach a peak at \(r \approx 4 r_0\)
(\(\approx \qty{0.3}{pc}\) or \qty{2.5}{arcmin}),
beyond which they fall again.
As discussed in \citet[][Sec.~3.2.3 and Fig.~4b]{garciav23},
this is an indication that the velocity fluctuations in the nebula are not homogeneous,
but instead show a decline in amplitude from the core to the outskirts.
As a result, all separations larger than half the size of the field must be between
two points in the outer regions, where the fluctuations are less vigorous.

An exception to this behavior is the unbinned \ariii{} line (Fig.~\ref{fig:Ha_Br_comparison}b),
for which the structure function initially falls after the \qty{0.3}{pc} peak,
but then rises again towards the largest separations.
As also shown in \citet[][Sec.~3.2.3]{garciav23}, this type of behavior can be evidence for
long-wavelength periodic fluctuations in the velocity field,
but this is unlikely to be the case here since the only significant difference
between the \ariii{} velocity map and the other lines (Fig.~\ref{fig:velocity_maps_shared_cbar_fixed})
is its lower signal-to-noise.
Indeed, it can be seen from Fig.~\ref{fig:Ha_Br_comparison}b that increased levels of binning
reduce and eventually eliminate the peak at large separations.
It is therefore likely to be caused by an inhomogeneous distribution of the
statistical uncertainty in the \(\vcentroid\) measurements:
the signal-to-noise is lowest in the fainter regions,
which tend to be towards the edges of the map,
as can be directly verified by inspecting Fig.~\ref{fig:velocity_maps_shared_cbar_fixed}.
As a result, the noise contribution to the structure function is highest for the largest separations,
which must be between two points at opposite edges.
In principal, it would be possible to exploit information in the surface brightness maps
to construct a more sophisticated noise model than the constant value that we assume in this paper.

\subsection{Comparison with previous works}

We compare our results with recent studies of the second-order structure function in the Orion Nebula, including \citet{2016MNRAS.455.4057M} and \citet{2019MNRAS.483..704A}, both of which analyzed the same VLT MUSE observations used here.

\subsubsection{Comparison with \citet{2016MNRAS.455.4057M}}\label{sec:comp_McLeod}

\citet{2016MNRAS.455.4057M} computed the second-order structure function following the method of \citet{2015MNRAS.447.1341B} for a randomly selected sample of \(10^3\) pixels \(j\), around which they radially bin all other pixels \(i\). 
They applied an intensity-based selection criterion for a mask, excluding pixels with \oi\ surface brightness below \qty{\sim 1e-16}{erg.s^{-1}.cm^{-2}.pixel^{-1}},
then computed the second-order structure function for both the masked and unmasked velocity maps, while also considering the influence of high-velocity objects.

In their analysis, the power-law index was estimated by fitting a straight line to the observed structure function over a restricted range (\(\approx \qty{0.7}{dex}\))
of spatial separations.
For the highest S/N lines of \halpha{} and \oiii{} they determined indices in the range
\(m = \numrange{0.22}{0.45}\),
which are significantly smaller than the value \(m = 1.1\) that we find for the same two lines (Table~\ref{tab:results_MUSE}).

They argue that the structure function cannot be reliably recovered due to lower spectral resolution, short exposure times and a low S/N ratio. 
However, as demonstrated in Section~\ref{sec:met}, the recovery of the structure function is strongly dependent on the analysis methodology, and appropriate treatment of noise and spatial resolution allows the turbulent signal to be recovered even in low signal-to-noise conditions.
Inspection of \citeauthor{2016MNRAS.455.4057M}'s measured \halpha{} structure function (their Fig.~12a)
reveals it to be similar to our own (our Fig.~\ref{fig:fit_results_01}a)
over the same spatial range, once the difference in normalizations is accounted for.
As recognized by \citet{2016MNRAS.455.4057M}, the shallow apparent slopes that they find are 
primarily due to not correcting for the effects of noise.

\subsubsection{Comparison with \citet{2019MNRAS.483..704A, Anorve-Zeferino:2023a}}
\label{sec:comp_Anorve}

In a pair of papers, \citet{2019MNRAS.483..704A, Anorve-Zeferino:2023a}
claim to apply a ``mathematically sound methodology'' to the calculation of velocity
structure functions, using the same MUSE \halpha{} velocity map from
\citet{2015A&A...582A.114W}.
However, the structure functions presented in those papers differ strongly
from those measured by us and by \citet{2016MNRAS.455.4057M} from the same observations.
We find that the structure function reaches a maximum at separations
\(\approx \qty{0.3}{pc}\) and then declines
(see section~\ref{sec:behavior-at-large}),
whereas \citet{2019MNRAS.483..704A} present structure functions in their Fig.~5
that increase monotonically before reaching a constant plateau at separations
\(> \qty{0.8}{pc}\).
Similar results are presented by \citet{Anorve-Zeferino:2023a},
where the computational method is described in greater detail
and extended to higher-order structure functions.
Unfortunately, the zero-padding employed in that method is fundamentally flawed:
it causes the resulting structure functions to be dominated by the boundary
of the valid data region and by the arbitrary choice of velocity zero point.

The large-scale decline in the correctly estimated Orion structure function
is not evidence of computational failure.
Instead, it reflects the spatial inhomogeneity of the velocity fluctuations,
together with the changing spatial distribution of valid pixel pairs as the
separation increases.
By padding the map with heliocentric zero velocities,
\citeauthor{2019MNRAS.483..704A} suppress this behaviour and replace it with
an artificial rise towards a normalization-imposed plateau.
For the second-order structure function, the normalization factor given by
eq.~(11) of \citet{Anorve-Zeferino:2023a} is proportional to
\(2\langle \vcentroid^2 \rangle\).
For the MUSE \halpha{} data, with \(\vcentroid\) measured in the heliocentric frame,
this is approximately
\[
2\langle \vcentroid^2 \rangle
=
2\big(
\langle \vcentroid \rangle^2 + \sigma^2
\big)
\approx
\qty{470}{km^2.s^{-2}}.
\]
This is much larger than the true fluctuation amplitude,
\(2\sigma^2 \approx \qty{18}{km^2.s^{-2}}\).
Consequently, the artificial contribution introduced by zero-padding overwhelms
the signal from the intrinsic velocity fluctuations in the
\citeauthor{Anorve-Zeferino:2023a} structure functions.

We also emphasize that the non-standard structure-function estimator
adopted by \citeauthor{Anorve-Zeferino:2023a} is not invariant under
a constant transformation of the velocity frame,
\[
\vcentroid(\boldsymbol{x})
\rightarrow
\vcentroid(\boldsymbol{x})+v_0.
\]
This lack of Galilean invariance makes it fundamentally unsuitable
for analyzing velocity fluctuations.

\subsubsection{Comparison with older studies}
\label{sec:comp-with-older}

There is a long history of studies of the velocity structure function in the Orion Nebula
\citep[e.g.,][]{castaneda1988, 1992ApJ...387..229O, Wen:1993a},
see discussion in Sec~4.1 of \citet{arthur2016turbulence}.
Although these earlier studies were limited in their spatial coverage,
they were based on spectra with higher velocity resolution than the more recent observations.
Table~5 of \citet{arthur2016turbulence} summarises the results for power-law slope and velocity dispersion
from the different studies, showing that both \(m\) and \(\sigma\pos^2\) tend to increase slightly
as a function of degree of ionization, in agreement with our results shown in Fig.~\ref{fig:parameters_comparison}.
In general, there is good consistency between different studies of the same ion,
despite the large differences in methodology.
The only outlier is the \oi{} line \citep{1992ApJ...387..229O},
which shows a significantly smaller plane-of-sky velocity variance (\(\sigma\pos^2 \approx \qty{3}{km^2.s^{-2}}\))
than other lines.
This line is of lower ionization than any of those considered in this paper,
and has its peak emissivity exactly at the ionization front \citep[Fig.~5c]{2021MNRAS.502.4597H}.
It may be that weakness of the velocity fluctuations seen in this line
is due to the fact that [\ion{O}{I}] traces the kinematics of the neutral gas of the
photodissociation region, rather than the ionized gas within the nebula.

\begingroup
\setlength{\tabcolsep}{6pt} 
\renewcommand{\arraystretch}{1.5} 
\begin{table*}
\begin{center}
  \caption{
    Best-fit model parameters and credibility intervals for fits to observed structure functions in the Orion core for the VLT MUSE observations.
    The results presented here correspond to binning level of $4\times4$ for all emission lines.
  }

  
\begin{tabular}{c RRR @{\hspace{6\tabcolsep}} RR @{\hspace{6\tabcolsep}} RRR}
\toprule
Emission
& \sigma^2\pos
& r_0
& m
& B_{\text{noise}}
& s_0 (\text{FWHM})
& \text{SNR}
& s_0 / r_0
& \text{Bin size} / r_0  \\
line
&[\si{km^2.s^{-2}}]
& [\si{pc}]
& [-]
& [\si{km^2.s^{-2}}]
& [\si{arsec}]
& [-]
& [-]
& [-] \\
\midrule
\n
&9.15\PM{0.41}{0.24} & 0.056\PM{0.004}{0.003} & 0.91\PM{0.02}{0.03} & 0.70\PM{0.15}{0.16} & <1.23 & 5.13 & <0.0176 & 0.027 \\ 

\siii
&10.00\PM{0.44}{0.33} & 0.065\PM{0.004}{0.003} & 1.10\PM{0.03}{0.03} & 0.64\PM{0.09}{0.06} & <1.09 & 5.60 & <0.0134 & 0.023\\

\ariii
&9.36\PM{0.51}{0.47} & 0.061\PM{0.005}{0.005} & 1.06\PM{0.05}{0.04} & 1.57\PM{0.19}{0.09} & <1.52 & 3.45 & <0.0200  & 0.025 \\

\halpha
& 8.88\PM{0.61}{0.50} & 0.068\PM{0.007}{0.005} & 1.16\PM{0.05}{0.04} & 0.55\PM{0.09}{0.05} & <1.23 & 5.69 & <0.0144 & 0.022 \\

\oiii
&8.49\PM{1.27}{0.94} & 0.068\PM{0.015}{0.011} & 1.11\PM{0.10}{0.08} & 0.85\PM{0.18}{0.10} & <1.72 & 4.48 & <0.0203 & 0.022 \\

\bottomrule

\end{tabular}\label{tab:results_MUSE}
\end{center}
\end{table*}
\endgroup

\begingroup
\setlength{\tabcolsep}{6pt} 
\renewcommand{\arraystretch}{1.5} 
\begin{table*}
\begin{center}
  \caption{
    Best-fit model parameters and credibility intervals for fits to observed structure functions in the Orion core for the KPNO echelle observations.
  }

  
\begin{tabular}{c RRR @{\hspace{6\tabcolsep}} RR @{\hspace{6\tabcolsep}} RRR}
\toprule
Emission
& \sigma^2\pos
& r_0
& m
& B_{\text{noise}}
& s_0 (\text{FWHM})
& \text{SNR}
& s_0 / r_0 \\
line
& [\si{km^2.s^{-2}}]
& [\si{pc}]
& [-]
& [\si{km^2.s^{-2}}]
& [\si{arcsec}]
& [-]
& [-] \\
\midrule
\sii
&7.03\PM{0.38}{0.38} & 0.044\PM{0.004}{0.003} & 1.01\PM{0.05}{0.05} & 0.062\PM{0.014}{0.016} & 1.91\PM{0.43}{0.39} & 15.04 & 0.0348 \\

\n
&8.15\PM{0.67}{0.53} & 0.037\PM{0.005}{0.003} & 1.04\PM{0.07}{0.09} & 0.034\PM{0.011}{0.010} & 2.13\PM{0.65}{0.44} & 22.04 & 0.0454  \\

\halpha
& 11.10\PM{0.53}{0.34} & 0.060\PM{0.004}{0.002} & 1.23\PM{0.02}{0.03} & 0.025\PM{0.004}{0.004} & 1.82\PM{0.23}{0.19} & 30.07 & 0.0244 \\

\oiii
& 10.48\PM{0.88}{0.56} & 0.058\PM{0.006}{0.004} & 1.23\PM{0.04}{0.05} & 0.022\PM{0.005}{0.007} & 1.74\PM{0.36}{0.35} & 31.03 & 0.0239 \\ 

\bottomrule

\end{tabular}\label{tab:results_KPNO}
\end{center}
\end{table*}
\endgroup

\begingroup
\setlength{\tabcolsep}{6pt} 
\renewcommand{\arraystretch}{1.5} 
\begin{table*}
\begin{center}
  \caption{
    Best-fit model parameters and credibility intervals for fits to observed structure functions in the Orion core for the VLT MUSE observations using KPNO echelle FoV (MUSE-trim).
    The results presented here correspond to binning level of $4\times4$ for all emission lines.
  }

  
\begin{tabular}{c RRR @{\hspace{6\tabcolsep}} RR @{\hspace{6\tabcolsep}} RRR}
\toprule
Emission
& \sigma^2\pos
& r_0
& m
& B_{\text{noise}}
& s_0 (\text{FWHM})
& \text{SNR}
& s_0 / r_0 \\
line
& [\si{km^2.s^{-2}}]
& [\si{pc}]
& [-]
& [\si{km^2.s^{-2}}]
& [\si{arcsec}]
& [-]
& [-] \\
\midrule

\n
&9.78\PM{0.82}{0.59} & 0.042\PM{0.007}{0.004} & 0.91\PM{0.04}{0.05} & 0.562\PM{0.334}{0.200} & <1.53 & 5.90 & <0.0290   \\

\halpha
&12.12\PM{1.38}{1.00} & 0.066\PM{0.010}{0.007} & 1.18\PM{0.05}{0.05} & 0.541\PM{0.117}{0.062} & <1.09 & 6.69 & <0.0133  \\

\oiii
&14.11\PM{4.31}{2.46} & 0.084\PM{0.036}{0.019} & 1.09\PM{0.09}{0.08} & 0.705\PM{0.182}{0.119} & <1.30 & 6.32 & <0.0124 \\

\bottomrule

\end{tabular}\label{tab:results_MUSE_trim}
\end{center}
\end{table*}
\endgroup

\section{Summary}\label{sec:summary}

Despite its relatively poor velocity resolution, intermediate
spectral-resolution integral-field spectroscopy can recover reliable
turbulent velocity statistics when analyzed with an appropriate
treatment of instrumental noise and spatial resolution. Applied to
the Orion Nebula, the VLT MUSE observations yield well-constrained
values of the turbulent velocity variance, \(\sigma^2\), correlation length, \(r_0\), and
power-law slope, \(m\), that agree closely with higher spectral-resolution
KPNO echelle observations. Our principal findings are as follows:

\begin{enumerate}[1.]

\item We have developed a methodology that reliably recovers the
turbulent velocity structure function from intermediate
spectral-resolution integral-field spectroscopy by combining spatial
masking, controlled binning, and model fitting.

\item Spatial binning provides an effective means of reducing
instrumental noise, provided that the bin size remains substantially
smaller than the turbulent correlation length. We find that reliable
recovery of the turbulent parameters is maintained for bin sizes
satisfying \(\Delta x_{\rm bin} \lesssim 0.05\,r_0\),
which provides a practical guideline for the analysis of similar
datasets.

\item Comparison with the KPNO echelle observations validates the
methodology across datasets with very different instrumental
characteristics, confirming that intermediate-resolution
integral-field spectroscopy can recover robust turbulent statistics
while providing substantially improved spatial sampling.

\item The different emission lines yield broadly similar turbulent
properties, with only modest systematic variations between ions.
Higher-ionization species tend to exhibit slightly larger velocity
variances, correlation lengths, and power-law indices, suggesting
subtle differences in the turbulent environments sampled by
different tracers.

\item Our analysis resolves previous discrepancies in structure
function studies of the Orion Nebula. In particular, we show that
apparent failures to recover the turbulent cascade arise primarily
from inadequate treatment of instrumental noise or from
methodological choices that introduce systematic biases into the
estimated structure function.

\end{enumerate}

\section*{Acknowledgements}
Based in part on observations made with the MUSE spectrograph
on VLT telescope UT4 at the La Silla Paranal Observatory, ESO, Chile.
Based in part on observations obtained at the Kitt Peak National Observatory,
which is operated by the Association of Universities for Research in Astronomy, Inc.,
under cooperative agreement with the National Science Foundation.
We gratefully acknowledge financial support from the \foreignlanguage{spanish}{Dirección General de Asuntos del Personal Académico (DGAPA), Universidad Nacional Autónoma de México (UNAM)}, through the ``\foreignlanguage{spanish}{Programa de Apoyo a Proyectos de Investigación e Innovación Tecnológica'' (PAPIIT), project IN117326}. J.G.-V. gratefully acknowledges the postdoctoral grant received through the ``\foreignlanguage{spanish}{Programa de Becas Posdoctorales en la Universidad Nacional Autónoma de México}'', which supported this work. A sincere thanks to Dr. C. Robert O'Dell who help us improve the quality of our work.

\section*{Data availability statement}
\label{sec:data-avail-stat}
The original MUSE datacube of the Orion Nebula is available at \url{https://data.aip.de/projects/musescience.html}.
Derived velocity maps and accompanying analysis programs used in this paper are currently available
from the GitHub repository \url{https://github.com/JavGVastro/VIENTO}
and from the Zenodo research repository at \url{https://doi.org/10.5281/zenodo.22261245}.



\bibliographystyle{mnras}
\bibliography{turb-refs}


\appendix

\section{Detailed results for all MUSE emission lines} \label{apex:multiple_lines_results}

\begin{figure*}
  \centering
  \fitfigggg{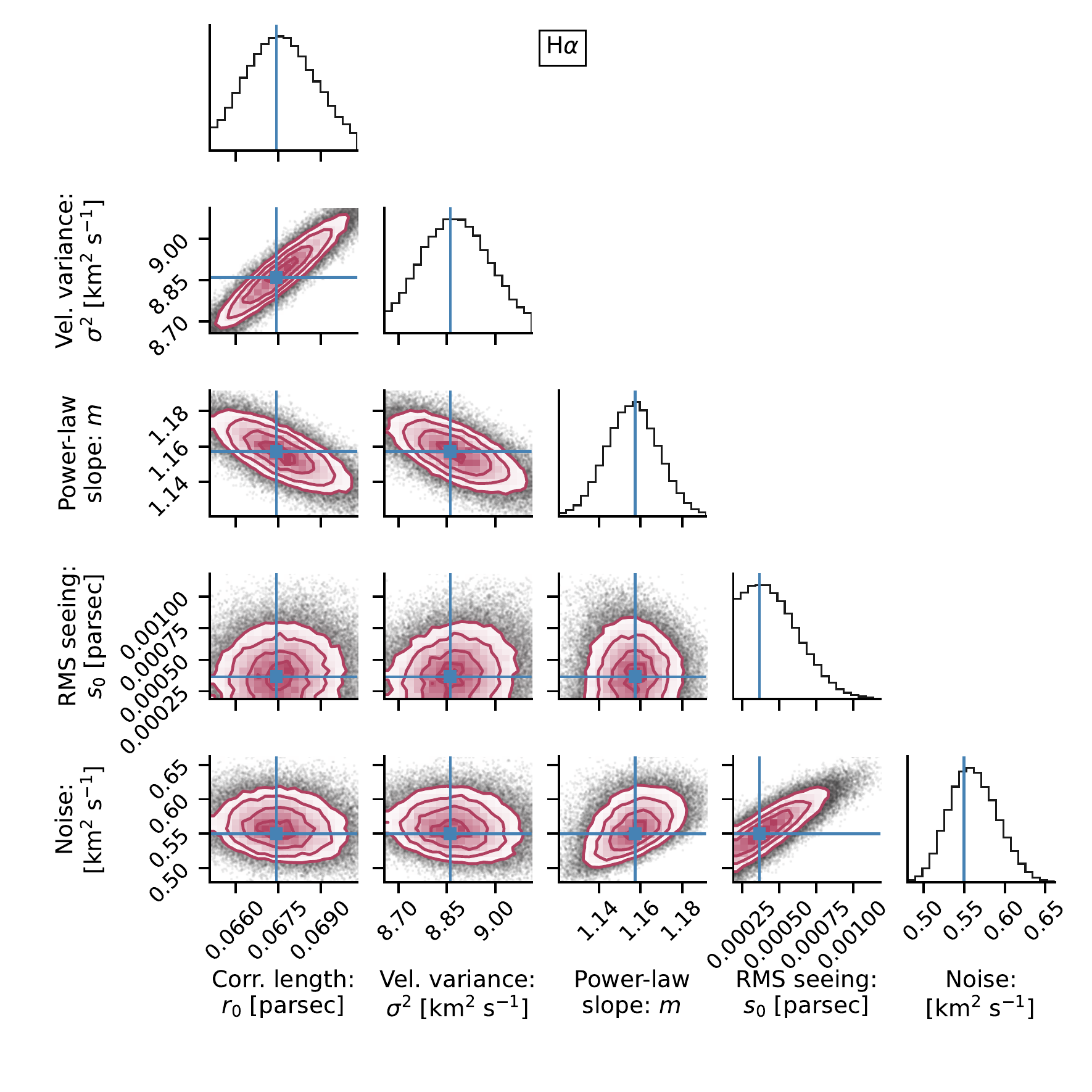}{MUSE-M42-H-bin2}{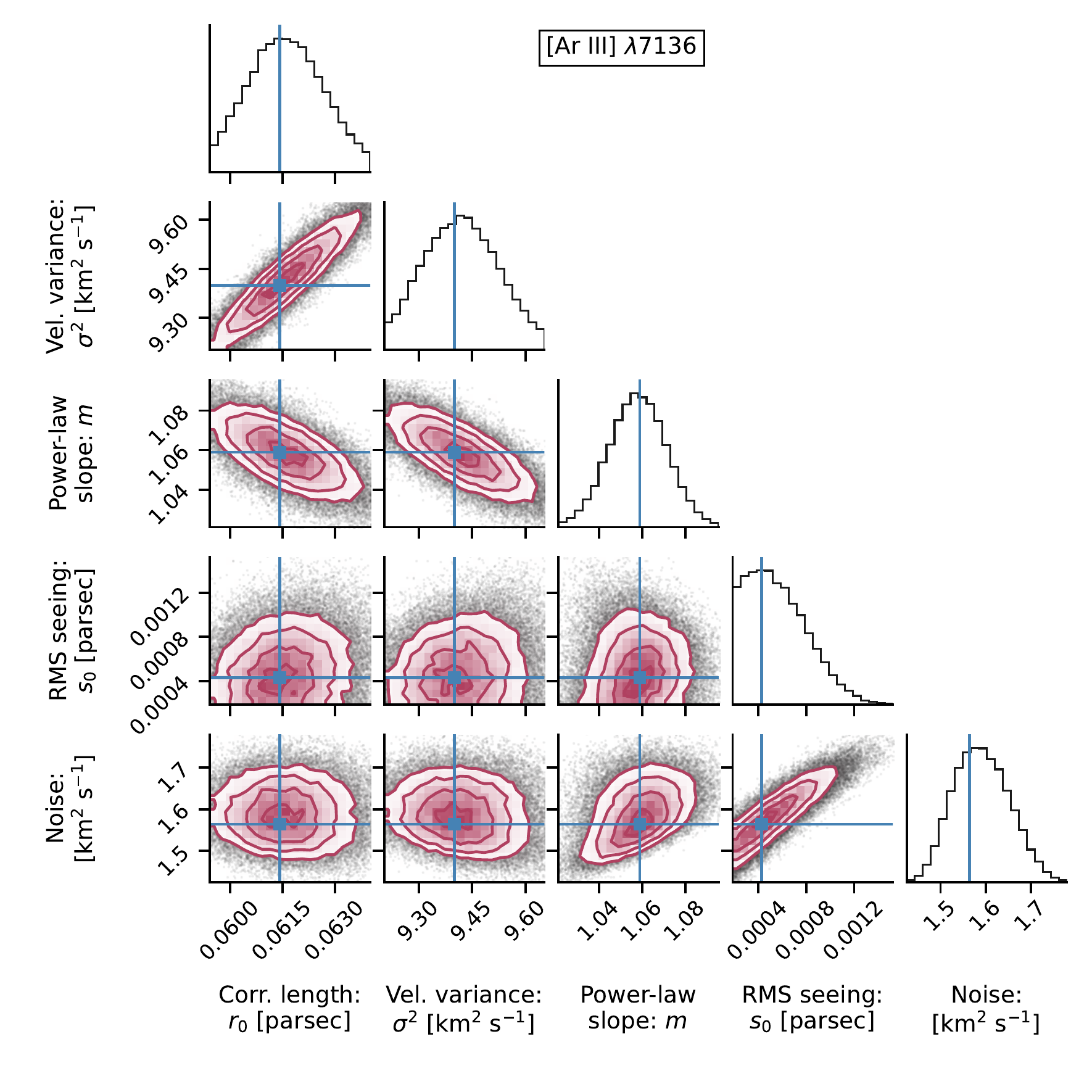}{MUSE-M42-Ar-bin2}
  \caption{ 
    (a)~Second-order structure functions
    for velocity centroid images
    of the \ha{} emission line from M~42 with a binning level of $4\times4$.
    The observed structure function \(B\obs(r)\) is shown by 
    blue symbols, with filled symbols indicating those points
    that are used to constrain the model fits.
    The full model is shown in orange,
    while the underlying model
    (without the effects of seeing or noise)
    is shown in green.
    See text for details of the fitting process.
    (b)~Corner plot of covariances between
    model parameters of fits to the \ha{} structure function.  Plots on the diagonal show the 1-dimensional histogram
    of the posterior distribution of each parameter
    (labeled at bottom),
    as calculated by the MCMC method,
    assuming a uniform prior distribution
    within the limits given in Table~\ref{tab:parameter-ranges}.
    Off-diagonal plots show the 2-dimensional histogram of the
    joint posterior distribution each pair of parameters.
    The results in the figures correspond to fits performed up to a maximum separation of \(0.5L\).
    (c)~Same as (a) for \ariii. (d)~Same as (b) for \ariii.
  }
  \label{fig:fit_results_01}
\end{figure*}

\begin{figure*}
  \centering
  \fitfigggg{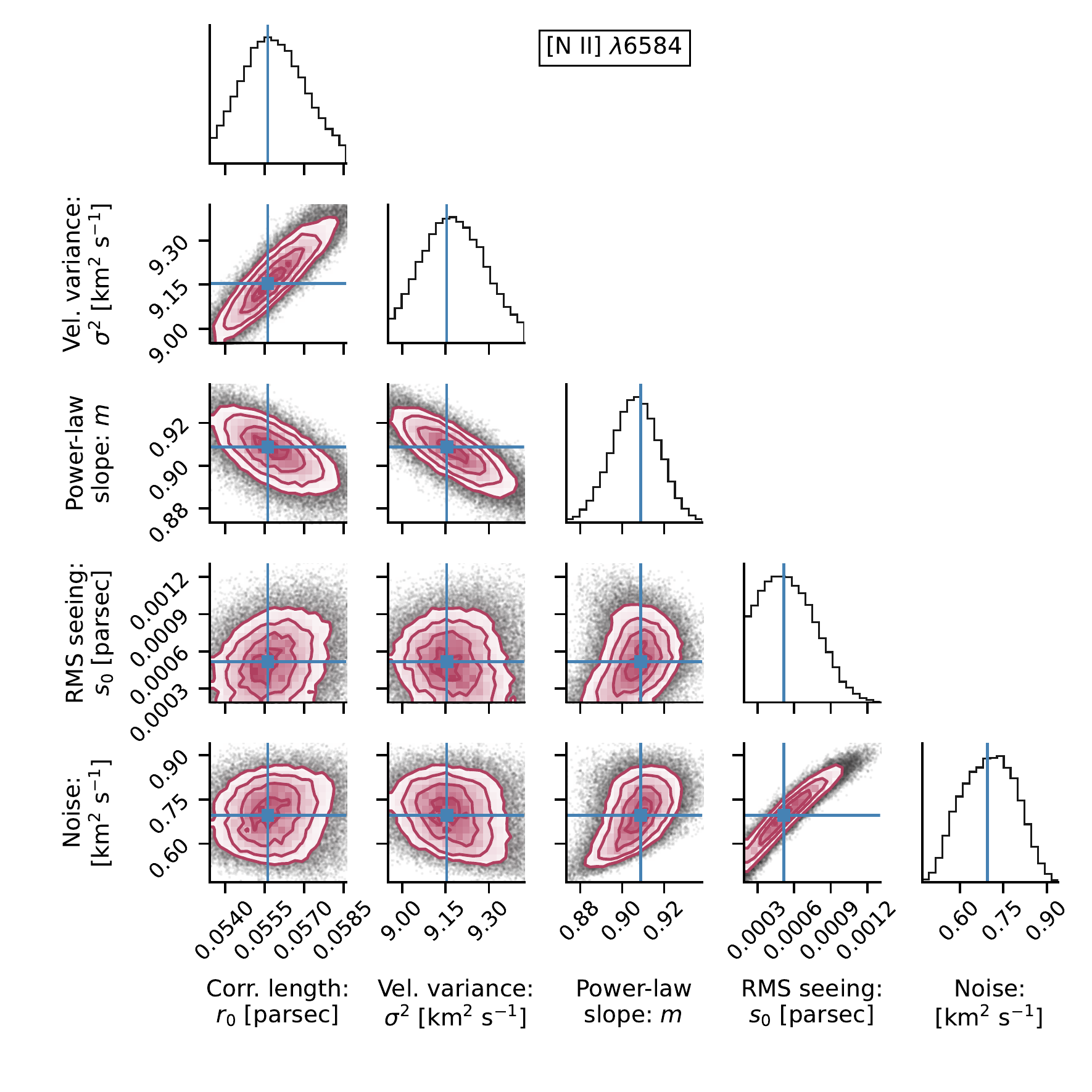}{MUSE-M42-N-bin2}{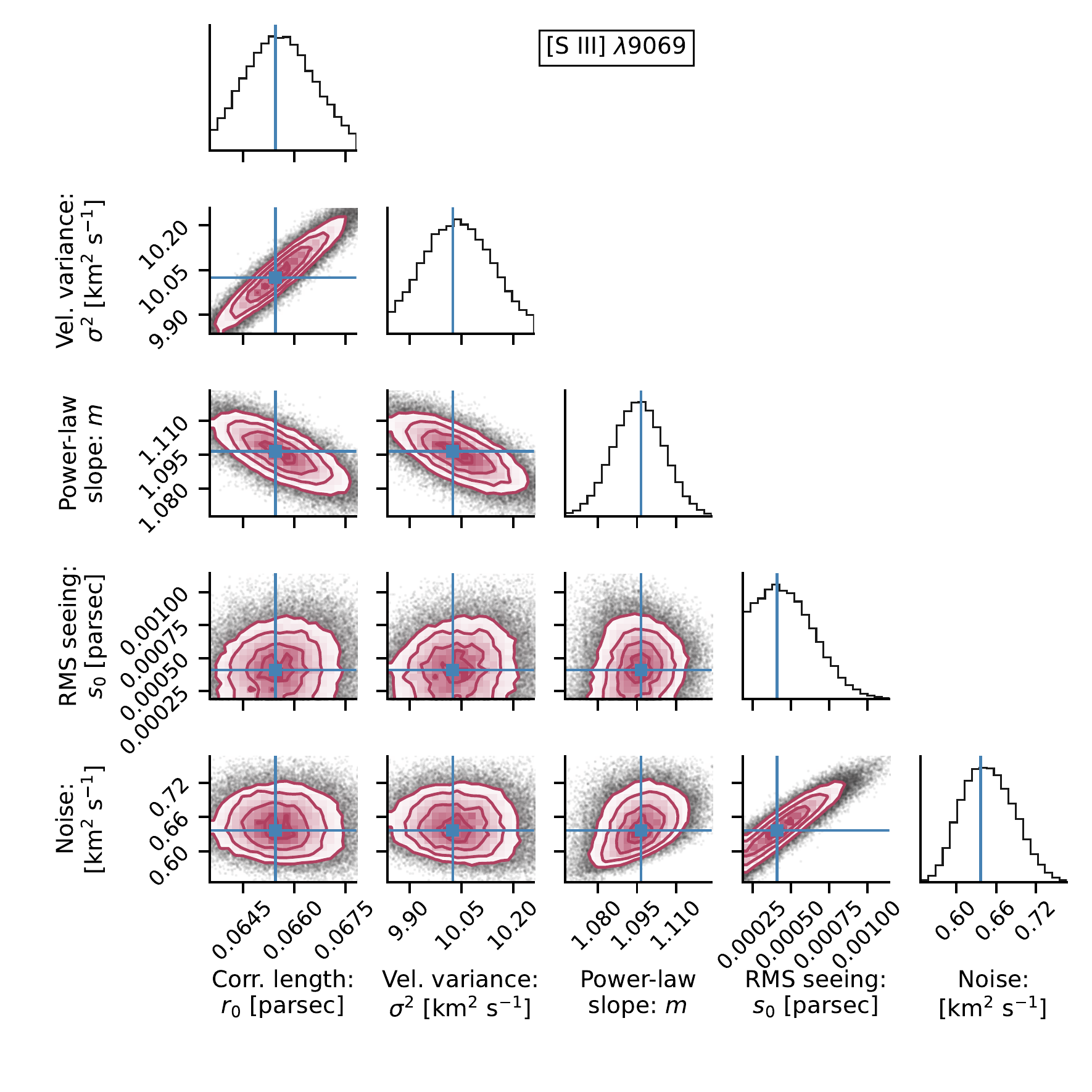}{MUSE-M42-S-bin2}
  \caption{ 
  Same information as Figure~\ref{fig:fit_results_01} for the \n\ and \siii\ emission lines obtained through VLT MUSE.
  }
  \label{fig:fit_results_02}
\end{figure*}

\begin{figure*}
  \centering
  \fitfigg{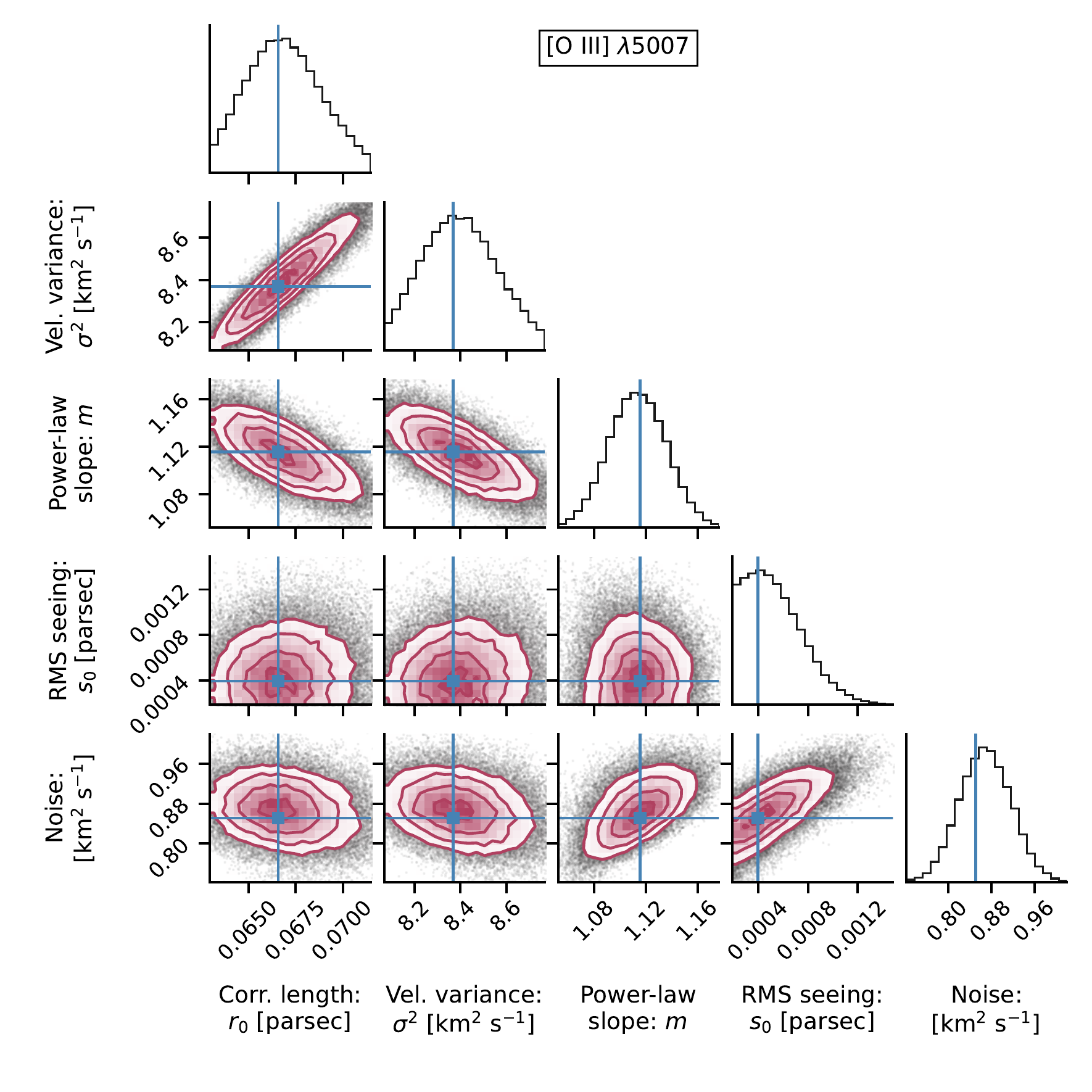}{MUSE-M42-O-bin2}
  \caption{  Same information as Figure~\ref{fig:fit_results_01} for the \oiii\ obtained through VLT MUSE.
  }
  \label{fig:fit_results_03}
\end{figure*}

This appendix presents additional tables and figures that supplement the results
given in the main text.
Tables~\ref{tab:results_MUSE_O}--\ref{tab:results_MUSE_S} list the
derived turbulent parameters and their confidence intervals for the additional emission lines
\oiii, \n, and \siii.
Figures~\ref{fig:fit_results_01}--\ref{fig:fit_results_03}
present the structure function fits plus
corner plots of the marginal posterior distributions of model parameters
for all emission lines from the  VLT MUSE observations.

\begingroup
\setlength{\tabcolsep}{6pt} 
\renewcommand{\arraystretch}{1.5} 
\begin{table*}
\begin{center}
  \caption{
    Best-fit model parameters and 95\% credibility intervals for fits to observed structure functions in the Orion core for the VLT MUSE \n\ line observations.
  }

  
\begin{tabular}{l RRR @{\hspace{6\tabcolsep}} RR @{\hspace{6\tabcolsep}} RRR}
\toprule
Binning
& \sigma^2\pos
& r_0
& m
& B_{\text{noise}}
& s_0 (\text{FWHM})
& \text{SNR}
& s_0 / r_0
& \text{Bin size} / r_0  \\
level
& [\si{km^2.s^{-2}}]
& [\si{pc}]
& [-]
& [\si{km^2.s^{-2}}]
& [\si{arcsec}]
& [-]
& [-]
& [-] \\
\midrule
1$\times$1
&9.22\PM{0.66}{0.38} & 0.058\PM{0.008}{0.004} & 0.88\PM{0.04}{0.04} & 3.948\PM{0.125}{0.069} & <0.52 & 2.16 & <0.0071 & 0.006 \\

2$\times$2
&9.17\PM{0.39}{0.24} & 0.056\PM{0.004}{0.003} & 0.90\PM{0.02}{0.03} & 1.598\PM{0.101}{0.091} & <0.78 & 3.39 & <0.0109 & 0.013 \\

4$\times$4
&9.15\PM{0.41}{0.24} & 0.056\PM{0.004}{0.003} & 0.91\PM{0.02}{0.03} & 0.696\PM{0.153}{0.163} & <1.23 & 5.13 & <0.0176 & 0.027 \\

8$\times$8
&9.05\PM{0.32}{0.18} & 0.055\PM{0.003}{0.002} & 0.94\PM{0.02}{0.03} & 0.405\PM{0.195}{0.384} & <2.75 & 6.69 & <0.0403 & 0.055 \\

16$\times$16
&8.98\PM{0.27}{0.37} & 0.054\PM{0.002}{0.003} & 0.96\PM{0.05}{0.04} & 0.291\PM{0.215}{0.291} & <3.53 & 7.85 & <0.0527 & 0.112 \\

\bottomrule

\end{tabular}\label{tab:results_MUSE_N}
\end{center}
\end{table*}
\endgroup
\begingroup
\setlength{\tabcolsep}{6pt} 
\renewcommand{\arraystretch}{1.5} 
\begin{table*}
\begin{center}
  \caption{
    Best-fit model parameters and 95\% credibility intervals for fits to observed structure functions in the Orion core for the VLT MUSE \siii\ line observations.
  }

  
\begin{tabular}{l RRR @{\hspace{6\tabcolsep}} RR @{\hspace{6\tabcolsep}} RRR}
\toprule
Binning
& \sigma^2\pos
& r_0
& m
& B_{\text{noise}}
& s_0 (\text{FWHM})
& \text{SNR}
& s_0 / r_0
& \text{Bin size} / r_0  \\
level
& [\si{km^2.s^{-2}}]
& [\si{pc}]
& [-]
& [\si{km^2.s^{-2}}]
& [\si{arcsec}]
& [-]
& [-]
& [-] \\
\midrule
1$\times$1
& 9.72\PM{0.61}{0.40} & 0.067\PM{0.006}{0.004} & 1.10\PM{0.04}{0.04} & 3.196\PM{0.088}{0.046} & <0.72 & 2.47 & <0.0085 & 0.006 \\

2$\times$2
&9.88\PM{0.44}{0.33} & 0.066\PM{0.004}{0.003} & 1.11\PM{0.03}{0.03} & 1.364\PM{0.069}{0.034} & <0.71 & 3.81 & <0.0087 & 0.011 \\

4$\times$4
&10.00\PM{0.44}{0.33} & 0.065\PM{0.004}{0.003} & 1.10\PM{0.03}{0.03} & 0.638\PM{0.090}{0.064} & <1.09 & 5.60 & <0.0134 & 0.023\\

8$\times$8
&10.06\PM{0.41}{0.29} & 0.063\PM{0.003}{0.003} & 1.11\PM{0.03}{0.03} & 0.366\PM{0.185}{0.145} & <2.43 & 7.41 & <0.0308 & 0.048\\

16$\times$16
& 9.99\PM{0.45}{0.28} & 0.062\PM{0.003}{0.002} & 1.13\PM{0.05}{0.04} & 0.274\PM{0.396}{0.202} & <3.46 & 8.54 & <0.0448 & 0.097 \\

\bottomrule

\end{tabular}\label{tab:results_MUSE_S}
\end{center}
\end{table*}
\endgroup
\begingroup
\setlength{\tabcolsep}{6pt} 
\renewcommand{\arraystretch}{1.5} 
\begin{table*}
\begin{center}
  \caption{
    Best-fit model parameters and 95\% credibility intervals for fits to observed structure functions in the Orion core for the VLT MUSE \oiii\ line observations.
  }

  
\begin{tabular}{l RRR @{\hspace{6\tabcolsep}} RR @{\hspace{6\tabcolsep}} RRR}
\toprule
Binning
& \sigma^2\pos
& r_0
& m
& B_{\text{noise}}
& s_0 (\text{FWHM})
& \text{SNR}
& s_0 / r_0
& \text{Bin size} / r_0  \\
level
& [\si{km^2.s^{-2}}]
& [\si{pc}]
& [-]
& [\si{km^2.s^{-2}}]
& [\si{arcsec}]
& [-]
& [-]
& [-] \\
\midrule
1$\times$1
&8.19\PM{1.44}{0.86} & 0.068\PM{0.018}{0.011} & 1.12\PM{0.12}{0.11} & 5.341\PM{0.193}{0.107} & <1.59 & 1.75 & <0.0185 & 0.006 \\

2$\times$2
&8.37\PM{1.14}{0.85} & 0.068\PM{0.013}{0.010} & 1.12\PM{0.10}{0.08} & 2.203\PM{0.167}{0.083} & <1.41 & 2.76 & <0.0166  & 0.011 \\

4$\times$4
&8.49\PM{1.27}{0.94} & 0.068\PM{0.015}{0.011} & 1.11\PM{0.10}{0.08} & 0.847\PM{0.188}{0.104} & <1.72 & 4.48 & <0.0203 & 0.022 \\

8$\times$8
&8.22\PM{0.93}{0.78} & 0.066\PM{0.010}{0.009} & 1.17\PM{0.11}{0.08} & 0.519\PM{0.260}{0.125} & <3.13 & 5.62 & <0.0381 & 0.046 \\

16$\times$16
&8.10\PM{0.85}{0.79} & 0.065\PM{0.008}{0.008} & 1.24\PM{0.15}{0.10} & 0.336\PM{0.305}{0.267} & <3.45 & 6.95 & <0.0423 & 0.092 \\

\bottomrule
\end{tabular}\label{tab:results_MUSE_O}
\end{center}
\end{table*}
\endgroup

\section{Structure functions for KPNO observations} \label{apex:kpno_structure_function}

In this appendix, we present the structure functions and corner plots (Figures~\ref{fig:fit_results_04}--\ref{fig:fit_results_05}) for the \halpha, \oiii, \n, and \sii\ emission lines for the KPNO echelle observations.

\begin{figure*}
 \centering
 \includegraphics[width=5in]{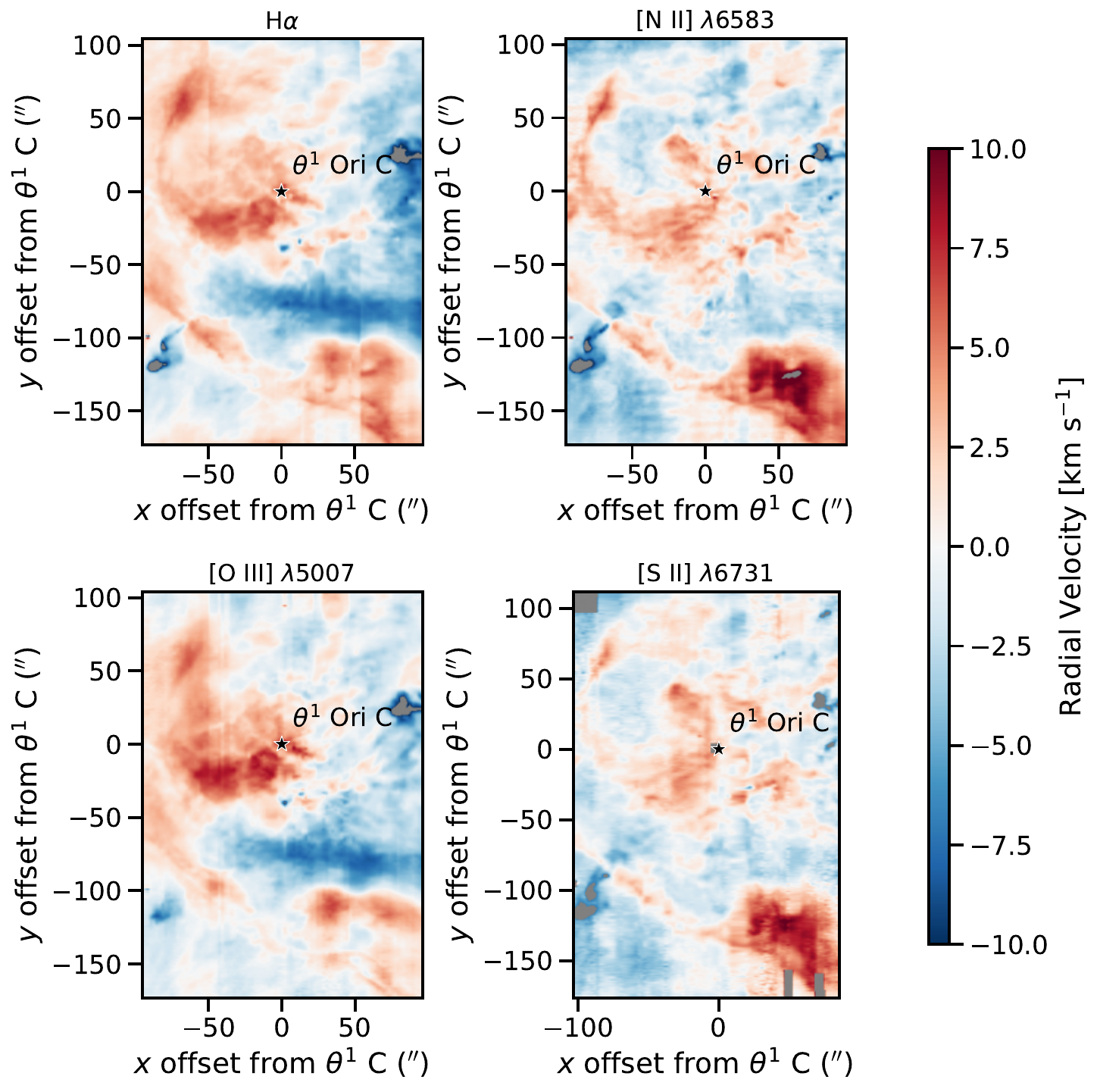}\par
 \caption{
 Two-dimensional KPNO velocity maps of the Orion Nebula in different emission lines. 
 North is up and east is to the left, with angular offsets measured relative to \(\theta^1\)~Ori~C. 
 High-velocity blueshifted features are masked. 
 The pixel scales are \SI{0.54}{arcsec\,pixel^{-1}} for \halpha, \n, and \oiii, and \SI{0.64}{arcsec\,pixel^{-1}} for \sii.
 }
\label{fig:kpno vel maps}
\end{figure*}

\begin{figure*}
  \centering
  \fitfigggg{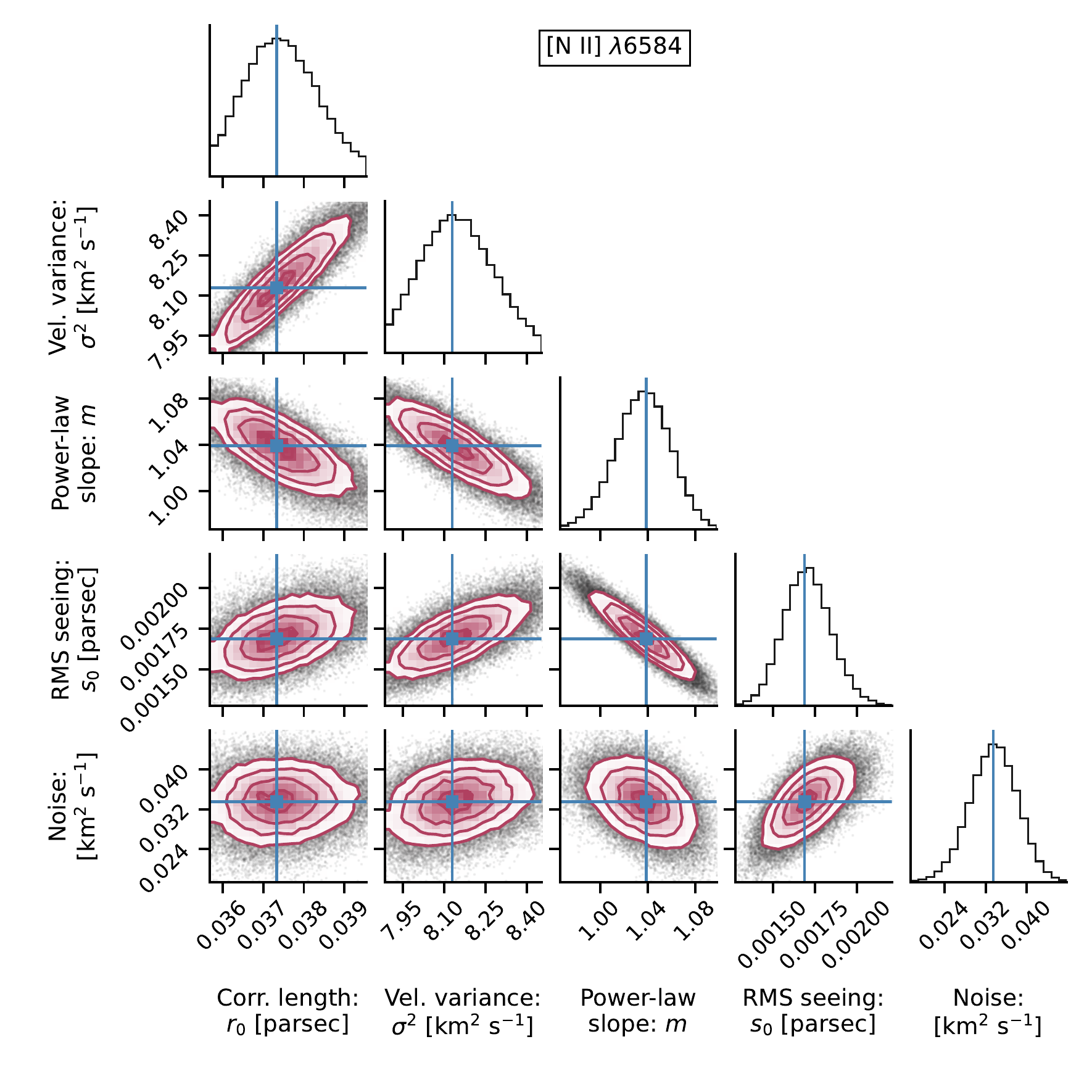}{KPNO-M42-N-bin0_05}{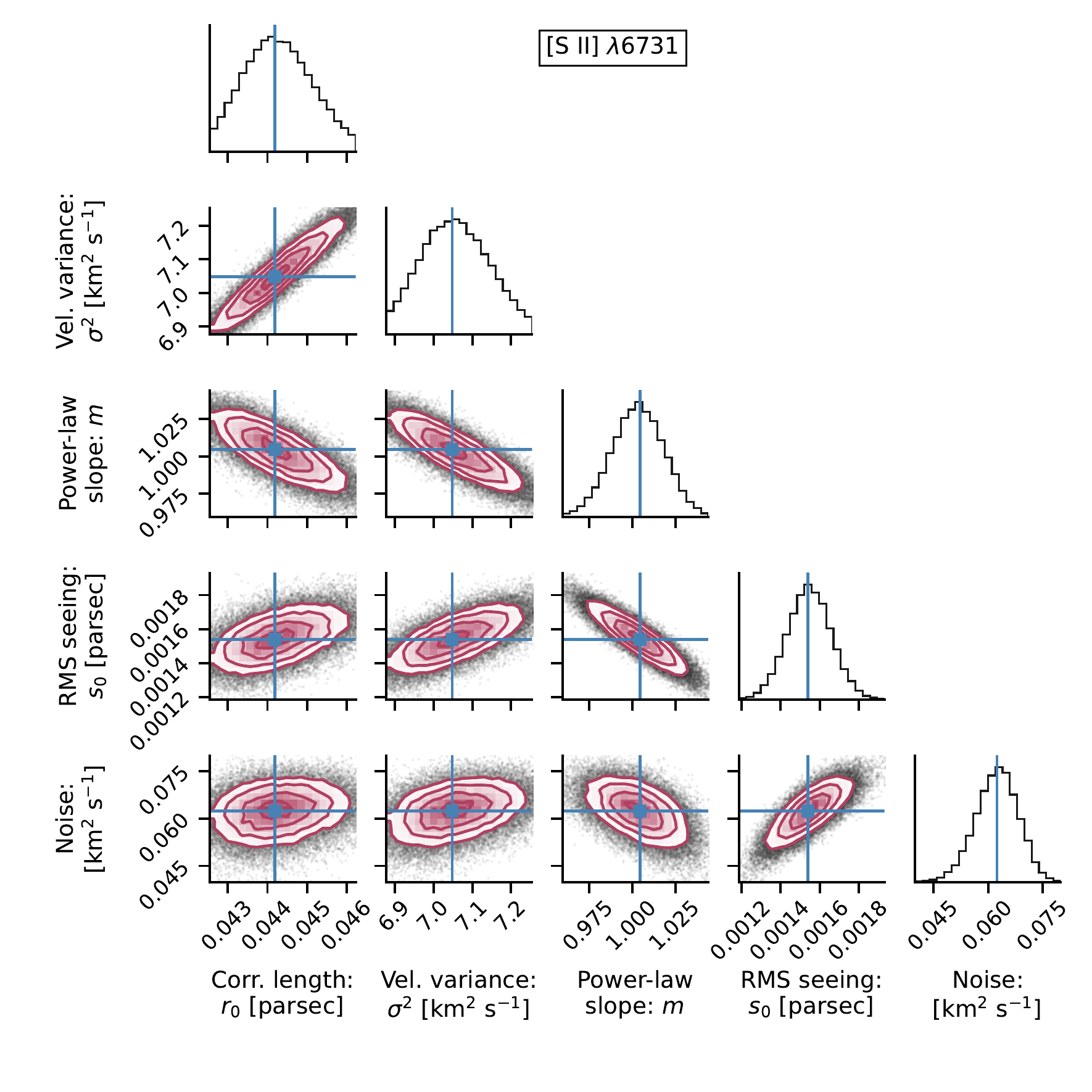}{KPNO-M42-S-bin0_05}
  \caption{  Same information as Figure~\ref{fig:fit_results_01} for the \n\ and \sii\ emission lines obtained through KPNO.
  }
  \label{fig:fit_results_05}
\end{figure*}

\begin{figure*}
  \centering
  \fitfigggg{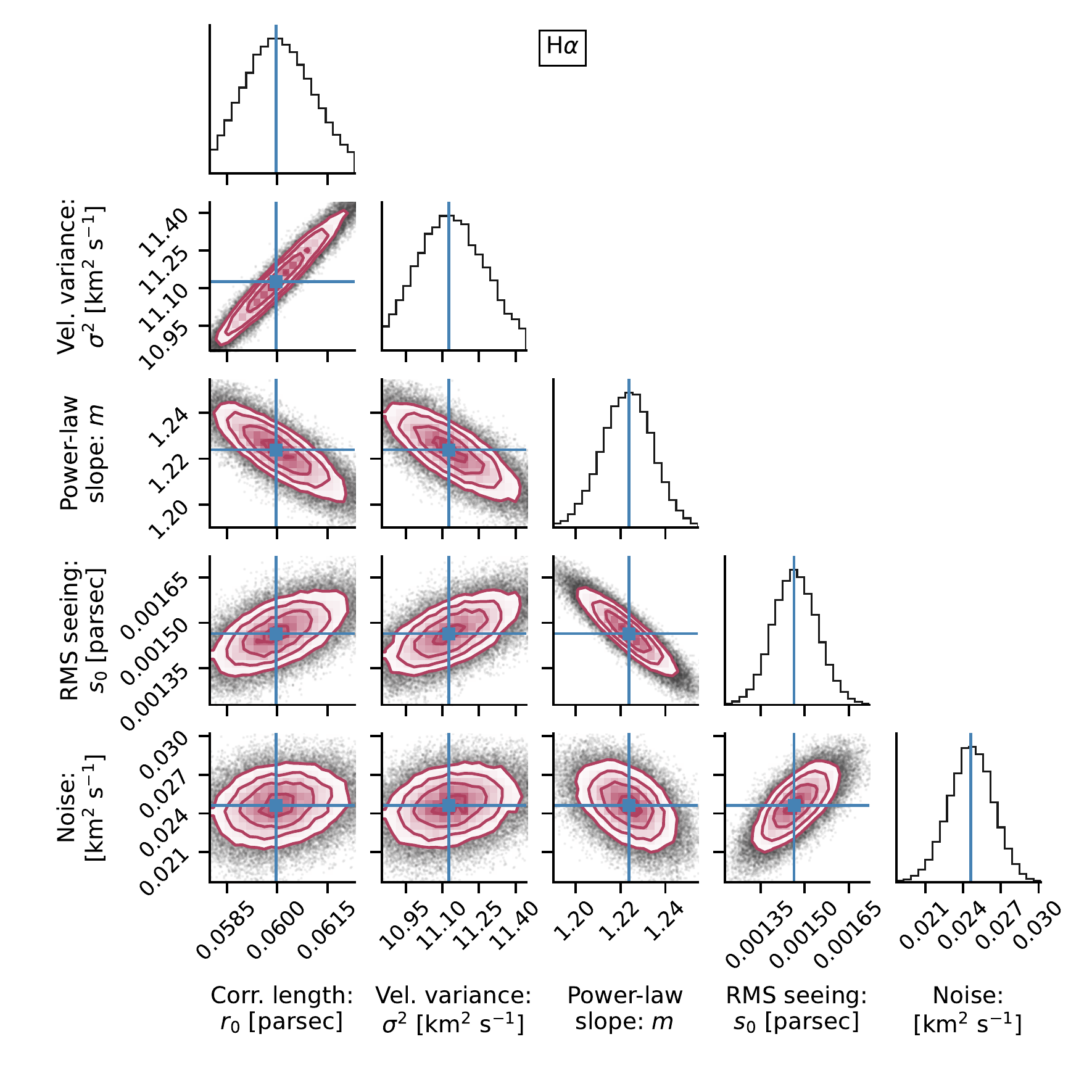}{KPNO-M42-H-bin0_05}{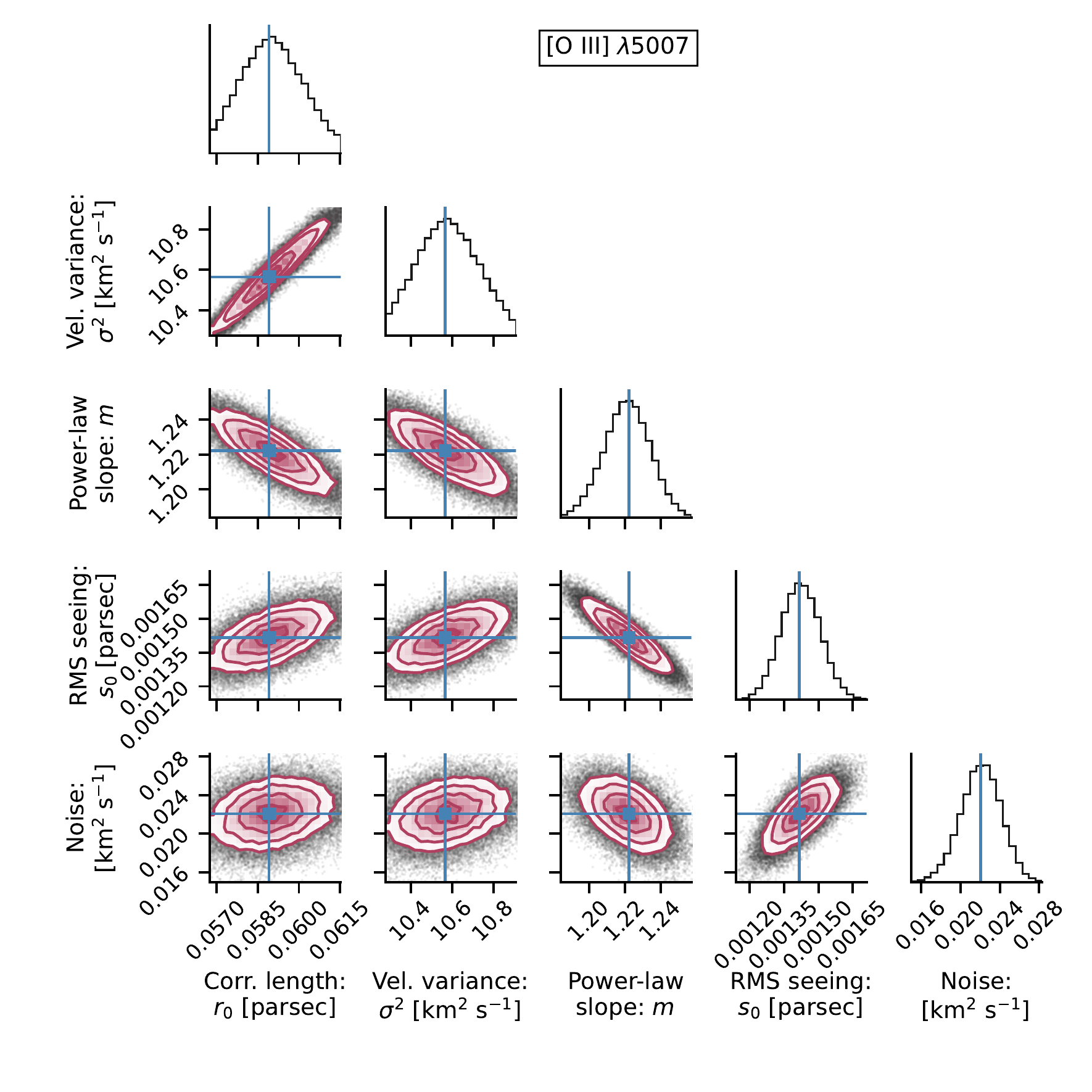}{KPNO-M42-O-bin0_05}
  \caption{ Same information as Figure~\ref{fig:fit_results_01} for the \halpha\ and \oiii\ emission lines obtained through KPNO.
  }
  \label{fig:fit_results_04}
\end{figure*}


%
\bsp	
\label{lastpage}
\end{document}